\documentclass[aps,prd,onecolumn,groupedaddress,nofootinbib,amssymb]{revtex4}
\usepackage{bm}

\usepackage{amsmath}
\usepackage{amssymb}
\usepackage{amsfonts}
\usepackage{wrapfig}
\usepackage{cancel} 
\usepackage[dvipdfmx]{graphicx}
\usepackage{color}
\newcommand{\be}{\begin{equation}}
\newcommand{\ee}{\end{equation}}
\newcommand{\bea}{\begin{eqnarray}}
\newcommand{\eea}{\end{eqnarray}}
\newcommand{\beaa}{\begin{eqnarray*}}
\newcommand{\eeaa}{\end{eqnarray*}}

\allowdisplaybreaks[4]

\begin{document}

\title{Bottom-up approach to massive spin-two theory in arbitrary curved spacetime}

\author{Satoshi Akagi$^{1}$\footnote{
	E-mail address: akagi.satoshi@b.mbox.nagoya-u.ac.jp }}

\affiliation{
$^1$ Department of Physics, Nagoya University, Nagoya
464-8602, Japan}

\begin{abstract}
The linear theory of massive spin-two field in arbitrary curved background is investigated. In  flat spacetime, the Fierz-Pauli model is well-known as the unique linear theory describing the massive spin-two field.
On the other hand, in order to construct the massive spin-two theory in fixed curved background with arbitrary metric, infinite series of nonminimal coupling terms are necessary.
In [Nucl.\ Phys.\ B {\bf 584} (2000) 615],  Buchbinder {\it et al.} have derived the condition for the ghost-freeness and they have solved the condition in small curvature approximation. In the leading order approximation, three free parameters of the leading order nonminimal coupling terms are allowed.
However, existence of the completion corresponding to all the three parameters 
is not guaranteed. 
On the other hand, recently, a class of the completion is obtained by linearizing the dRGT model. However, the leading order nonminimal coupling terms of the linearized dRGT model depend only on one free parameter. 
Therefore, possibility of a class larger than the linearized dRGT model has not been excluded.
In this paper, we develop the method for solving the conditions for ghost freeness in higher order and investigate whether lower order nonminimal coupling terms can be constrained by higher order conditions or not. As a result, we obtain an additional constraint on the leading order nonminimal coupling terms from the fourth order condition.
Although the leading order nonminimal coupling terms of the linearized dRGT model is still a subclass of our resulting nonminimal coupling terms in the leading order, a trivial extension of the linearized dRGT model is perfectly equivalent to our resulting action.
\end{abstract}

%\pacs{}

\maketitle
\section{INTRODUCTION}
\label{int}
\subsection{Bottom-up approach}
In flat spacetime, the linear theory of massive spin-two field has been well-known since a long time ago. It is called the Fierz-Pauli model \cite{Fierz:1939ix}.
On the other hand, in fixed curved background with arbitrary metric, it is well-known that nonminimal coupling terms are necessary in order to construct the ghost-free model.
This is because the minimal coupled model contains a ghost-like scalar mode in addition to massive spin-two mode due to the violation of a constraint.
Bubinder {\it et al.} have derived the condition for ghost-freeness and solved the condition in the leading order approximation with respect to $R/m^2$
(Here, $R$ denotes symbolical curvature and $m^2$ denotes the graviton mass squared).
The resulting action satisfying the leading order condition is given by \cite{Buchbinder1},\footnote{The difference between our notation and the notation of \cite{Buchbinder1} is summarized in Appendix \ref{ap2}.}
\if0
\begin{align}
S= \int d^Dx \sqrt{-g} &\left[ 3 \nabla_{\mu_1} h_{\mu_2}^{~[\mu_2 } \nabla^{\mu_1} h_{\mu_3}^{~\mu_3]}  
+\frac{1}{2} \left\{ 2m^2 g^{\mu_1 [\nu_1}g^{ \nu_2] \mu_2} 
+ 2\gamma_1 R g^{\mu_1 [\nu_1}g^{ \nu_2] \mu_2} \right. \right. \notag \\
 &\left. \left.+\frac{\gamma_2}{2} \left(R^{\mu_1 [\nu_1}g^{ \nu_2] \mu_2} - R^{\mu_2 [\nu_1} g^{\nu_2]  \mu_1} \right) 
+ \gamma_3 R^{\mu_1 \mu_2 \nu_1 \nu_2}\right\} h_{\mu_1 \nu_1} h_{\mu_2 \nu_2}+ \mathcal{O}\left(R^2/m^2 \right) \right]. 
\end{align}
\fi
\begin{align}
S= \int d^Dx \sqrt{-g} &\left[ \frac12 g^{\mu_1 \nu_1 \mu_2 \nu_2 \mu_3 \nu_3}\nabla_{\mu_1} h_{\mu_2 \nu_2 } \nabla_{\nu_1} h_{\mu_3 \nu_3}  
+\frac{1}{2} \left\{ m^2 g^{\mu_1 \nu_1 \mu_2 \nu_2} 
+ \gamma_1 R g^{\mu_1 \nu_1 \mu_2 \nu_2} \right. \right. \notag \\
 &\left. \left.+ \frac{\gamma_2}{2} \left(R^{\mu_1 [\nu_1}g^{ \nu_2] \mu_2} - R^{\mu_2 [\nu_1} g^{\nu_2]  \mu_1} \right) 
+ \gamma_3 R^{\mu_1 \mu_2 \nu_1 \nu_2}\right\} h_{\mu_1 \nu_1} h_{\mu_2 \nu_2}+ \mathcal{O}\left(R^2/m^2 \right) \right].\label{int1}
\end{align}
Here, although the detailed definition of the higher rank tensor $g^{\mu_1 \nu_1 \mu_2 \nu_2 \cdots \mu_n \nu_n}$ is given in the end of this section,
 $ g^{\mu_1 \nu_1 \mu_2 \nu_2 \mu_3 \nu_3}\nabla_{\mu_1} h_{\mu_2 \nu_2 } \nabla_{\nu_1} h_{\mu_3 \nu_3}$
denotes the kinetic term in our notation, 
$m^2 g^{\mu_1 \nu_1 \mu_2 \nu_2} h_{\mu_1 \nu_1} h_{\mu_2 \nu_2}$ denotes the Fierz-Pauli mass term, and the other terms denote the nonminimal coupling terms.
The higher order terms $\mathcal{O}(R^2/m^2)$ cannot be truncated for the ghost-freeness but cannot be determined by the leading order condition.
We find that this action contains three free parameters
$\gamma_1, \gamma_2, \gamma_3$.
However, it is NOT guaranteed that there is full completion of nonminimal coupling terms corresponding to all the above parameter region.
There remains possibility that the above free parameters may be constrained by higher order conditions.

\subsection{Historical introduction to massive gravity}
On the other hand, recently, a class of the full completion is obtained \cite{bernard1} from  metric perturbation of the de Rham-Gabadadze-Tolley (dRGT) model \cite{deRham:2010ik,deRham:2010kj}. The dRGT model is gravitational model whose metric perturbation has  finite graviton mass. 
Historically, it had been involved by many problems to regard the massive spin-two field as the gravity consisting with the observational results. In particular, although it had become necessary to non-linearize the Fierz-Pauli model in order to explain the observational results in the solar system \cite{vanDam:1970vg,Vainshtein:1972sx}, it had been pointed out that large class of the nonlinearization of the Firez-Pauli theory contains an additional degrees of freedom with negative kinetic terms \cite{Boulware:1974sr}. The additional mode appearing by non-linearizing the Fiertz-Pauli model is called the Boulware-Deser ghost. Although the ghost of the linear theory in the fixed curved background is not commonly called the BD ghost, we should note that the appearance of both ghosts is based on same reason, that is the violation of the constraint.
The non-linear model had not been constructed during a long time, because the requirement of the BD ghost-freeness is quite strong. 

However, in 2011, the BD ghost problem had been solved by discovering the dRGT model. 
At first, dRGT model was derived to fulfill the following assumptions: The kinetic term is chosen as the Einstein-Hilbert action, and the non-derivative potential terms are some functions of the metric $g_{\mu\nu}$ and the ``finite" fluctuation of the metric from the flat metric $h_{\mu\nu} \equiv g_{\mu\nu} - \eta_{\mu\nu}$. Under the assumption, requiring the ghost-freeness in the high energy limit called the decoupling limit, the first dRGT theory was derived \cite{deRham:2010ik,deRham:2010kj}. Then the resulting action depends on the dynamical metric $g_{\mu\nu}$ and the fixed flat metric $\eta_{\mu\nu}$.
Thereafter, it had been confirmed that appropriate constraints exist in the full theory without taking any limit \cite{Hassan:2011hr,Hassan:2011ea}.\footnote{Nowadays, the existence of the constraint is confirmed by several works \cite{kluson1,kluson2,kluson3,kluson4,golovnev,mirbabayi,HassanS,comelli,kugo1}.}
Furthermore, it was pointed out that the model replaced the flat metric $\eta_{\mu\nu}$ with  arbitrary fiducial metric $f_{\mu\nu}$ is also ghost-free \cite{Hassan:2011tf}.
This fact link the dRGT model to the linear theory of the massive spin-two field in arbitrary curved background.

\subsection{Complete nonminmal coupling terms from dRGT model}
Recently, some class of the complete linear theory of the massive spin-two field in arbitrary background is found by linearizing the dRGT model \cite{bernard1}. We would like to explain a simple idea of \cite{bernard1}, and derive the most general nonminimal coupling terms in the leading order with respect to curvature.
The action of the dRGT model is given by,
\begin{align}
&S_{\text{dRGT}} [g;f]= M_g^{D-2} \int d^D x\sqrt{-g}  \left[ R(g) 
-2m^2\sum_{n=0}^{D-1} \beta_n e_n (\mathcal{S}) \right], \notag \\
&e_n(\mathcal{S}) \equiv  
\mathcal{S}^{[\mu_1}_{ \ \ \mu_1} \mathcal{S}^{\mu_2}_{ \ \ \mu_2} \cdots \mathcal{S}^{\mu_n ]}_{ \ \ \mu_n} , \ \ 
\mathcal{S}^\mu_{ \ \ \nu} \equiv {\sqrt{g^{-1}f}}^\mu_{ \ \ \nu }
 , \ \ \mathcal{S}^\mu_{ \ \ \nu} \mathcal{S}^{\nu}_{ \ \ \rho} = g^{\mu \nu}f_{\nu \rho}.
\label{int2}
\end{align}
Here the matrix $\mathcal{S}^\mu_{~\nu}$ is square root of matrix $g^{\mu \nu} f_{\nu \rho}$, the matrix $f_{\mu \nu}$ is an arbitrary non-dynamical metric called the fiducial metric, $M_g$ is the Plank mass, $\beta_n$ are dimensionless parameters, and $m^2$ is a mass parameter introduced just in order to make $\beta_n$ dimensionless.\footnote{A choice of the free parameters is obtained from following requirements (a), (b):
(a)There are no tadpoles around the flat space when we choose the fiducial metric to be $f_{\mu\nu}=\eta_{\mu\nu}$. (b)The mass parameter $m^2$ becomes the Fierz-Pauli mass in the linearized equation. In Appendix \ref{Ldr}, we show how we choose the free parameters.}

The metric perturbation of the action (\ref{int2}) around the on-shell background is ghost-free because the full theory has appropriate constraints \cite{Hassan:2011hr,Hassan:2011ea,kluson1,kluson2,kluson3,kluson4,golovnev,mirbabayi,HassanS,comelli,kugo1,Hassan:2011tf}. If we restrict the fiducial metric to be $f_{\mu\nu}=\eta_{\mu\nu}$, we obtain a linear theory which is ghost-free only around the solutions of the
EoM of the dRGT theory with $f_{\mu\nu}=\eta_{\mu\nu}$. However, if we consider the set of the metric perturbation of the dRGT theories corresponding to various fiducial metrics, this set describes the massive spin-two theory in fixed curved spacetime with arbitrary metric. 
In fact, for any metric $g_{\mu\nu}$, we can determine the fiducial metric $f_{\mu\nu}(g)$ so that the metric $g_{\mu\nu}$ is a solution of the EoM with the fiducial metric $f_{\mu\nu}(g)$. In order to concretize this idea, let us note that the action (\ref{int2}) does not contains any derivatives of the fiducial metric $f_{\mu\nu}$. 
Then, we can solve algebraically the EoM with respect to $f_{\mu\nu}$ as some functions of the metric $g_{\mu\nu}$. We would like to denote the solution as $f_{\mu\nu}(g)$. Therefore, we can obtain the linear theory in arbitrary background by substituting the algebraic solution $f_{\mu\nu} = f_{\mu\nu}(g)$ into the linearized action of dRGT theory (\ref{int2}). Performing this procedure, in the leading order, we obtain,
\begin{align}
S_{\text{dRGT}}&\left[g +\frac{\sqrt{2}}{M_g^{(D-2)/2}}h  \ ; \ f(g)\right]_{\text{linear}}\notag \\
=&\int d^Dx \sqrt{-g} \left[  \frac12 g^{\mu_1 \nu_1 \mu_2 \nu_2 \mu_3 \nu_3}\nabla_{\mu_1} h_{\mu_2 \nu_2 } \nabla_{\nu_1} h_{\mu_3 \nu_3}  
+\frac{1}{2} \left\{ m^2 g^{\mu_1 \nu_1 \mu_2 \nu_2} 
+\frac{s_2D -1}{2(D-1)}  R g^{\mu_1 \nu_1\mu_2 \nu_2} \right. \right. \notag \\
 &\left. \left.- 2s_2\left(R^{\mu_1 [\nu_1}g^{ \nu_2] \mu_2} - R^{\mu_2 [\nu_1} g^{\nu_2]  \mu_1} \right) 
+  R^{\mu_1 \mu_2 \nu_1 \nu_2}\right\} h_{\mu_1 \nu_1} h_{\mu_2 \nu_2}+ \mathcal{O}\left(R^2/m^2 \right) \right].\label{int3}
\end{align}
Here the parameter $s_2$ is a function of $\beta_n$ in (\ref{int2}) and can be regarded as a free parameter.
\footnote{The explicit form of $s_2$ and the derivation of the action (\ref{int3}) is summarized in Appendix \ref{Ldr}.
Furthermore, similar result has been obtained in \cite{fukuma}.}
We find that the linearized dRGT model (\ref{int3}) is certainly included in the bottom-up result (\ref{int1}) by tuning the free parameters as follows,
\begin{align}
 \gamma_1=\frac{s_2D -1}{2(D-1)}, \ \ \gamma_2=-4s_2, \ \ \gamma_3=1. \label{pp1}
\end{align}
However, we find that the linearized dRGT model depends only on one free parameter $s_2$. 
In contrast to the bottom-up case (\ref{int1}), it is guaranteed that there is full completion of the nonminimal coupling terms corresponding to all the above parameter region (\ref{pp1}). 

\subsection{Our purpose}
Comparing the bottom-up result (\ref{int1}) and the linearized dRGT model (\ref{pp1}), it seems that there may be a class larger than the linearized dRGT model.
In fact, we have proposed that the infinite number of the free parameters are allowed in the case of the Einetein manifold \cite{new curved}.
\if0
The resulting nonminimal coupling terms of \cite{new curved} depend on Weyl curvatures in addition to scalar curvatures. 
In the leading order, our result is just equivalent to the Einstein manifold limit of Buchbinder's result in (\ref{int1}).
Thus, in the case of Einstein manifold, the free parameter $\gamma_3$ are not constrained in contrast to the tuning of the linearized dRGT model, $\gamma_3=1$.
\fi
However, in the case of the arbitrary background, there remains a possibility that the free parameters in (\ref{int1}) may be constrained by higher order conditions.
Hence, in this paper, we develop the method to solve the higher order conditions for the ghost-freeness, and investigate whether the three free parameters in Eq.(\ref{int1}) are constrained by the higher order conditions or not.
As a result, from the fourth order condition, we obtain a constraint on the three free parameters in the leading order nonminimal coupling terms.
Although the leading order nonminimal coupling terms of the linearized dRGT model is still a subclass of our resulting nonminimal coupling terms in the leading order, a trivial extension of the linearized dRGT model is perfectly equivalent to our resulting action.

This paper is organized as follows: In Sec.\ref{lag}, we introduce the method for counting the degrees of freedom in Lagrangian formulation by using the Fierz-Pauli model in flat space.
In Sec.\ref{irr}, we give an assumption and introduce a simple formulation which makes the calculation quite easy.
In Sec.\ref{cond}, we apply the Lagrangian analysis introduced in Sec.\ref{lag} to the case of curved background and derive a condition which makes the theory ghost-free.
In Sec.\ref{pert}, we solve the condition obtained in Sec.\ref{cond} perturbatively and derive the most general nonminimal coupling terms consisting with the condition in order by order.
As a result of this section, we obtain an additional constraint on the free parameters in (\ref{int1}). 
In Sec.\ref{cdr}, restricting our argument in the leading order nonminimal coupling terms, we compare our result obtained in Sec.\ref{pert} with the linearized dRGT model.
Although the linearized dRGT model is still a subclass of our resulting action,
we will see that there is a trivial extension of the linearized dRGT model whose leading order nonminimal coupling terms are perfectly equivalent to our resulting nonminimal coupling terms.

Furthermore, in Appendix \ref{ap1} and \ref{ap2},
the notation used in this paper is summarized.
In Appendix \ref{Ldr}, we give the detailed derivation of the linearized dRGT model (\ref{int3}).
The case of the Einstein manifold is argued in Appendix \ref{ein} although we will do not mention it in the main parts. 
In the case of the Einstein manifold, we will see that the condition given in Sec.\ref{cond} can be solved in full-order.
Finally, in the Appendix \ref{str}, the relationships between our approach and the String theory is argued.

\subsection{Notation}
In this paper, we use the higher rank tensor $\delta^{\mu_1~~\mu_2~~ \cdots \mu_n }_{~~~\nu_1~~\nu_2 ~~\cdots \nu_n}$ defined as the anti-symmetrization of the product $\delta^{\mu_1}_{\nu_1} \delta^{\mu_2}_{\nu_2} \cdots \delta^{\mu_n}_{\nu_n}$ with respect to $\nu_1 \nu_2 \cdots \nu_n$, i.e., 
\begin{align}
\delta^{\mu_1~~\mu_2~~ \cdots \mu_n }_{~~~\nu_1~~\nu_2 ~~\cdots \nu_n}
\equiv n! \delta^{\mu_1}_{[\nu_1} \delta^{\mu_2}_{\nu_2} \cdots \delta^{\mu_n}_{\nu_n]}.
\end{align}
Furthermore, we define the higher rank tensor $g^{\mu_1 \nu_1 \mu_2 \nu_2 \cdots \mu_n \nu_n}$ as follows,
\begin{align}
g^{\mu_1 \nu_1 \mu_2 \nu_2 \cdots \mu_n \nu_n}
\equiv \delta^{\mu_1~~\mu_2~~ \cdots \mu_n }_{~~~\rho_1~~\rho_2 ~~\cdots \rho_n}
g^{\rho_1\nu_1} g^{\rho_2 \nu_2} \cdots g^{\rho_n \nu_n}.
\end{align}
A few examples are given by,
\begin{align}
g^{\mu_1 \nu_1 \mu_2 \nu_2} \equiv& g^{\mu_1 \nu_1}g^{ \mu_2 \nu_2} - g^{\mu_1 \nu_2}g^{\mu_2 \nu_1}, \notag \\
g^{\mu_1 \nu_1 \mu_2 \nu_2 \mu_3 \nu_3} \equiv& 
g^{\mu_1 \nu_1} g^{\mu_2 \nu_2} g^{\mu_3 \nu_3}
+ g^{\mu_1 \nu_2} g^{\mu_2 \nu_3} g^{\mu_3 \nu_1}
+g^{\mu_1 \nu_3} g^{\mu_2 \nu_1} g^{\mu_3 \nu_2}\notag \\
&-g^{\mu_1 \nu_2} g^{\mu_2 \nu_1} g^{\mu_3 \nu_3}
-g^{\mu_1 \nu_1} g^{\mu_2 \nu_3} g^{\mu_3 \nu_2}
-g^{\mu_1 \nu_3} g^{\mu_2 \nu_2} g^{\mu_3 \nu_1}. \label{aaa1}
\end{align}
We should note that the superscripts are automatically anti-symmetrized with respect to $\mu_i$ in addition to $\nu_i$. 
We lower the indexes of $g^{\mu_1 \nu_1 \mu_2 \nu_2 \cdots \mu_n \nu_n}$ by using the metric $g_{\mu\nu}$ and raise the indexes by using the inverse matrix $g^{\mu\nu}$.
The detailed properties of this tensor is summarized in Appendix \ref{ap1}.
As the same way, we define $\eta^{\mu_1 \nu_1 \mu_2 \nu_2 \cdots \mu_n \nu_n}$
and $\delta^{i_1 j_1 i_2 j_2 \cdots i_n j_n}$ in the flat spacetime,
\begin{align}
&\eta^{\mu_1 \nu_1 \mu_2 \nu_2 \cdots \mu_n \nu_n}
\equiv \delta^{\mu_1~~\mu_2~~ \cdots \mu_n }_{~~~\rho_1~~\rho_2 ~~\cdots \rho_n}
\eta^{\rho_1\nu_1} \eta^{\rho_2 \nu_2} \cdots \eta^{\rho_n \nu_n}, \notag \\
&\delta^{i_1 j_1 i_2 j_2 \cdots i_n j_n}
\equiv \delta^{i_1~~i_2~~ \cdots i_n }_{~~~k_1~~k_2 ~~\cdots k_n}
\delta^{k_1j_1} \delta^{k_2 j_2} \cdots \delta^{k_n j_n}.
\end{align}
By using these notations, the Fierz-Pauli action in the flat spacetime can be expressed as follows,\footnote{For example, this notation is used in \cite{Hinterbichler:2013eza}.}
\begin{align}
S_{\mathrm{FP}} &= \int d^Dx \left[-\frac{1}{2}\partial_\lambda h_{\mu\nu}\partial^\lambda 
h^{\mu\nu}+\partial_\mu h_{\nu\lambda}\partial^\nu h^{\mu\lambda}-\partial_\mu 
h^{\mu\nu}\partial_\nu h+\frac{1}{2}\partial_\lambda h\partial^\lambda h 
 -\frac{1}{2}m^2(h_{\mu \nu}h^{\mu \nu}-h^2)\right] \notag \\
 &=\int d^Dx \left[\frac12 \eta^{\mu_1 \nu_1 \mu_2 \nu_2 \mu_3 \nu_3} \partial_{\mu_1} h_{\mu_2 \nu_2} \partial_{\nu_1} h_{\mu_3 \nu_3}
 + \frac{m^2}{2}\eta^{\mu_1 \nu_1 \mu_2 \nu_2} h_{\mu_1 \nu_1} h_{\mu_2 \nu_2}\right].
\end{align}

\section{LAGRANGIAN ANALYSIS}
\label{lag}
In this paper, we count the degrees of freedom (DoF) by using the Lagrangian formulation. Such method is called the Lagrangian analysis.
We would like to introduce Lagrangian analysis by counting DoF of the Fierz-Pauli model in the flat spacetime.

The Fierz-Pauli theory is described by the following EoM,
\begin{align}
E^{\mu \nu} \equiv -\eta^{ \mu \nu  \mu_1 \nu_1 \mu_2 \nu_2} \partial_{\mu_1} 
\partial_{\nu_1} h_{\mu_2 \nu_2} + m^2 \eta^{\mu \nu  \mu_1 \nu_1}h_{\mu_1 \nu_1} =0. \label{f1}
\end{align}
Here, we omit the symmetrization factor $()$ from the superscripts $\mu\nu$ of the higher rank tensors $\eta^{\mu \nu \mu_1 \nu_1 \cdots \mu_n \nu_n}$ because these superscripts are automatically symmetrized. 
Let us investigate the dependence on the second order time derivative $\ddot{h}_{\mu\nu}$ in  Eq.(\ref{f1}),
\footnote{
On the first term of Eq.(\ref{f1}), if we take $(\mu_1, \nu_1) =(0,0)$, the other superscripts cannot be taken as $0$ because of the antisymmetry of the higher rank tensor $\eta^{\mu_1 \nu_1 \mu_2 \nu_2 \mu_3 \nu_3}$.  
The explicit form of Eq.(\ref{f5}) can be derived by using the following expansion,
\begin{align}
\eta^{\mu_1 \nu_1 \mu_2 \nu_2 \mu_3 \nu_3} 
= \eta^{\mu_1 \nu_1 } \eta^{\mu_2 \nu_2 \mu_3 \nu_3} + \eta^{\mu_1 \nu_2 } \eta^{\mu_2 \nu_3 \mu_3 \nu_1} 
+ \eta^{\mu_1 \nu_3} \eta^{\mu_2 \nu_1 \mu_3 \nu_2}.
\end{align}
The derivation of the above identity is given in (\ref{Ap3}).}
\begin{align}
&E^{ij} = \delta^{ij i_1 j_1} \ddot{h}_{i_1 j_1} + (\text{terms without } \ddot{h})^{ij} =0, \label{f5} \\
&\phi^{(1)\nu} \equiv E^{0\nu} = -\eta^{0 \nu  i_1 \nu_1 i_2 \nu_2} \partial_{i_1} \partial_{\nu_1} 
h_{i_2 \nu_2} + m^2 \eta^{0 \nu i_1 \nu_1}h_{i_1 \nu_1}=0. \label{f2}
\end{align}
First, we find that the spatial components (\ref{f5}) contain the second time derivative of  $h_{ij}$. The other terms of Eq.(\ref{f5}) do not contain any second time derivative terms due to the anti-symmetry of the sixth rank tensor $\eta^{\mu_1 \nu_1 \mu_2 \nu_2 \mu_3 \nu_3}$. Then, we can solve Eq.(\ref{f5}) for the quantity $\ddot{h}_{ij}$ as some function of the lower order time derivative terms. On the other hand, the $(0\mu)$ components of the EoM (\ref{f2}) do not contain any second time derivative terms. Then we find that the second time derivative of $h_{0\mu}$ does not appear in any components of the EoM (\ref{f1}).
Let us consider to determine the acceleration $\ddot{h}_{0\mu}$ by taking the time derivative of Eq.(\ref{f2}).

The time derivative of Eq.(\ref{f2}) is given by,
\begin{align}
\dot{\phi}^{(1)\nu} =0. \label{nf1}
\end{align}
Eq.(\ref{nf1}) guarantees the conservation of the quantity $\phi^{(1)\nu}$ for the time evolution. Hence, if Eq.(\ref{f2}) is only satisfied at the initial time, Eq.(\ref{f2}) is valid for any time. Then, we can regard the Eq.(\ref{f2}) as the ``constraint" on the initial values. In order to emphasize this fact, we would like to introduce the equal $\approx$ which means the equivalence at the initial time. By using the equal $\approx$, Eq.(\ref{f2}) is rewritten as follows,
\begin{align}
\phi^{(1)\nu} \approx 0. \label{nf2}
\end{align}
Furthermore, we would like to call the equation obtained by taking the time derivative of the constraint, such as Eq.(\ref{nf1}), the ``consistency condition". 
We continue the above procedure until the acceleration $\ddot{h}_{0\mu}$ appears in the consistency condition. By counting the number of the constraints obtained through this procedure, we can determine the DoF.

Let us determine the acceleration $\ddot{h}_{0\mu}$.
By adding some linear combinations of the EoM or the constraints, Eq.(\ref{nf1}) can be deformed as follows,
\footnote{The divergence of the kinetic term becomes equal to zero due to the anti-symmetry of the higher rank tensor $\eta^{\mu_1 \nu_1 \mu_2 \nu_2 \mu_3 \nu_3}$ and commutativity of the partial derivatives.}
\begin{align}
0 = \dot{\phi}^{(1)\nu} \approx \partial_\mu E^{\mu \nu} = m^2 \eta^{\mu \nu  \mu_1 \nu_1} \partial_\mu 
h_{\mu_1 \nu_1}  \equiv \phi^{(2)\nu}. \label{f6}
\end{align}
Then, we regard again Eq.(\ref{f6}) as constraint $\phi^{(2)\nu}\approx 0$ and require the consistency condition $\dot{\phi}^{(2)\nu}=0$, 
\footnote{Eq.(\ref{nana2}) is easily derived by using the identity (\ref{Ap2}) in Appendix \ref{ap1}.}
\begin{align}
&0=\dot{\phi}^{(2)i} = m^2 \delta^{ij} \ddot{h}_{0j} + (\text{terms without }\ddot{h})^{i},\label{nana1}  \\
&0 = \dot{\phi}^{(2)0}
\approx \partial_\mu \partial_\nu E^{\mu \nu} + \frac{m^2}{D-2} 
\eta^{\mu \nu}E_{\mu \nu}  
=  \frac{D-1}{D-2}m^4 h  \equiv \phi^{(3)}.\label{nana2}
\end{align}
From Eq.(\ref{nana1}), we can determine the acceleration $\ddot{h}_{0i}$
On the other hand, (\ref{nana2}) can be regarded as the constraint and we require the consistency condition,
\begin{align}
\phi^{(4)}\equiv \dot{\phi}^{(3)} =0. \label{nana3}
\end{align} 
This equation is also constraint.
The consistency condition of (\ref{nana3}), $\dot{\phi}^{(4)} =0$, contains the acceleration $\ddot{h}_{00}$. Then, we have completed the Lagrangian analysis.
Finaly, we obtain the $2(D+1)$ constraints $\phi^{(1)\nu}\approx 0,\phi^{(2)\nu}\approx 0,\phi^{(3)}\approx 0,\phi^{(4)}\approx 0$ between the $D(D+1)$ initial values.
Therefore, the Fierz-Pauli model has the $(D+1)(D-2)/2$ degrees of freedom.

\section{IRREDUCIBLE DECOMPOSITION}
\label{irr}
Let us consider the curved background case.
It is difficult to calculate the higher order nonminimal coupling terms with giving its explicit forms from the beginning, 
because the possible nonminimal coupling terms are too many in the higher order. 
In this section, we give a simple formulation which makes the calculation quite easy.

In this paper, we assume that the nonminimal coupling terms do not contain any derivatives acting on the massive spin-two field.
In other words, we assume the action expressed as follows,
\begin{align}
S=\int d^Dx \sqrt{-g} \left[\frac12 g^{\mu_1 \nu_1 \mu_2 \nu_2 \mu_3 \nu_3}\nabla_{\mu_1} h_{\mu_2 \nu_2 } \nabla_{\nu_1} h_{\mu_3 \nu_3}  
+\frac{1}{2} \left\{ m^2 g^{\mu_1 \nu_1 \mu_2 \nu_2} +U^{\mu_1 \nu_1 \mu_2 \nu_2}\right\}
h_{\mu_1 \nu_1} h_{\mu_2 \nu_2}\right]. \label{nana6}
\end{align}
Here, $U^{\mu_1 \nu_1 \mu_2 \nu_2}$ is an arbitrary covariant tensor, without the covariant derivative acting on the massive spin-two field $h_{\mu \nu}$, depending only on the first or higher order terms with respect to curvatures. 
We should note that the kinetic terms defined in the above action (\ref{nana6}) is not equivalent to common kinetic terms of the linearized Einstein-Hilbert action.
The difference is summarized in Appendix \ref{ap2}.

Next, let us consider the formulation which makes the analysis easy.
We reconsider Buchbinder's result in (\ref{int1}).
The bases of the nonminimal coupling terms which can be added to the action are given by,
\begin{align}
R g^{\mu_1 \nu_1 \mu_2 \nu_2}, \ \  R^{\mu_1 [\nu_1}g^{ \nu_2] \mu_2} - R^{\mu_2 [\nu_1} g^{\nu_2]  \mu_1} , \ \ R^{\mu_1 \mu_2 \nu_1 \nu_2}. \label{nana4}
\end{align}
We find that all the three bases have the same symmetry as the tensor $T^{\mu_1 \nu_1 \mu_2 \nu_2}$ satisfying the following relation,
\begin{align}
&T^{\mu_1 \nu_1 \mu_2 \nu_2} = - T^{\mu_2 \nu_1 \mu_1 \nu_2} = - T^{\mu_1 \nu_2 \mu_2 \nu_1} = T^{\mu_2 \nu_2 \mu_1 \nu_1}, \notag \\
&T^{[\mu_1 \nu_1 \mu_2] \nu_2}=0.\label{nana5}
\end{align}
\begin{table}
\centering
\includegraphics[width=13cm]{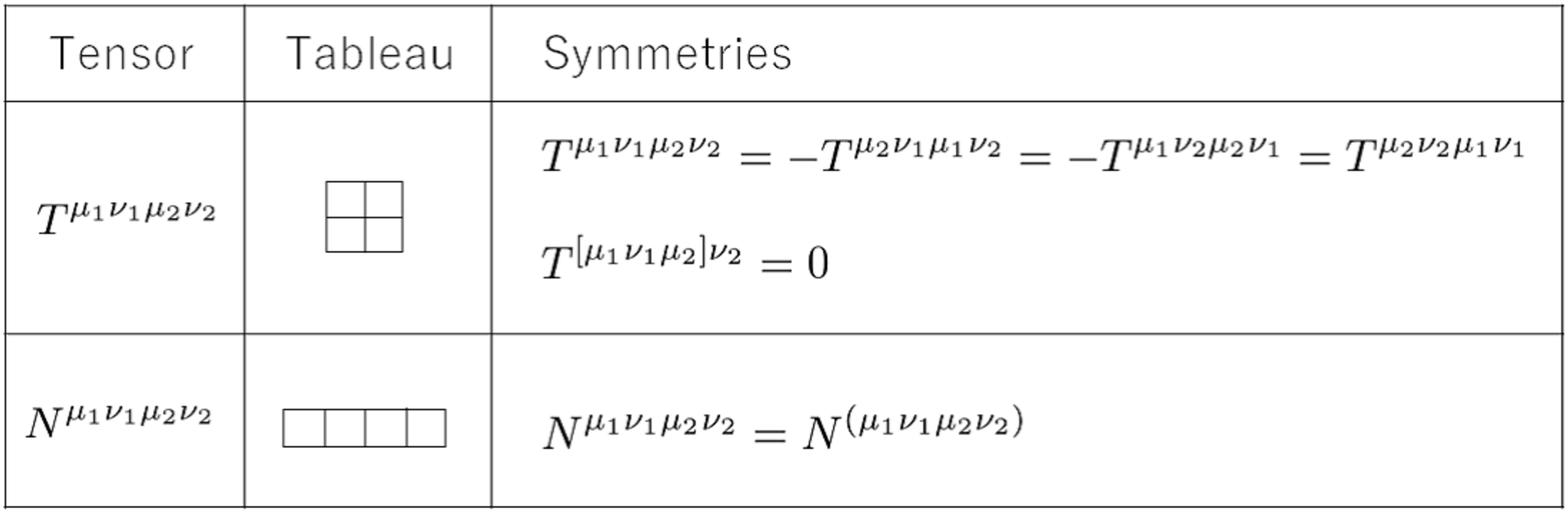}
\caption{Young tableau}
\label{fig1}
\end{table}
This symmetries correspond to the mixed symmetric tensor whose Young tableau is expressed in TABLE \ref{fig1}. Furthermore, the terms in (\ref{nana4}) are all the possible terms which have the mixed symmetry (\ref{nana5}) in the leading order. In other words, the nonminimal coupling terms constructed by the mixed symmetric tensor are not constrained by the leading order condition. We can easily predict that the explicit form of the leading terms (\ref{nana4}) are not necessary to solve the leading order condition, but these symmetries (\ref{nana5}) are necessary.
Then, we would like to decompose the general nonminimal coupling terms in (\ref{nana6}) into the terms expressed by the irreducible tensors.
We can show the following identity,
\begin{align}
&U^{\mu_1 \nu_1 \mu_2 \nu_2}h_{\mu_1 \nu_1}h_{\mu_2 \nu_2}
= T^{\mu_1 \nu_1 \mu_2 \nu_2}h_{\mu_1 \nu_1}h_{\mu_2 \nu_2}
+N^{\mu_1 \nu_1 \mu_2 \nu_2}h_{\mu_1 \nu_1}h_{\mu_2 \nu_2}, \notag \\
&T^{\mu_1 \nu_1 \mu_2 \nu_2} \equiv \frac12 \left[ T_0^{\mu_1 \nu_1 \mu_2 \nu_2} +  T_0^{\mu_2 \nu_2 \mu_1 \nu_1} \right], \notag \\ 
&T_0^{\mu_1 \nu_1 \mu_2 \nu_2} \equiv \frac{1}{2}\left( U^{[\mu_1| \nu_1 |\mu_2] \nu_2}
-U^{[\mu_1| \nu_2 |\mu_2] \nu_1} \right) + \frac16\left( U^{\mu_1 [\nu_1 \nu_2] \mu_2}
-U^{\mu_2 [\nu_1 \nu_2] \mu_1} \right) 
+\frac13 U^{ [\nu_2 \nu_1] [\mu_1 \mu_2]},\notag \\
&N^{\mu_1 \nu_1 \mu_2 \nu_2}\equiv U^{(\mu_1 \nu_1 \mu_2 \nu_2)}.\label{tn1}
\end{align}
for any tensor $U^{\mu_1 \nu_1 \mu_2 \nu_2}$, by a straightforward calculation. 
We find that the tensor $T^{\mu_1 \nu_1 \mu_2 \nu_2}$ has the mixed symmetry (\ref{nana5}), and the tensor $N^{\mu_1 \nu_1 \mu_2 \nu_2}$ is totally symmetric tensor.
Therefore, it is general even if we assume that the nonminimal coupling terms are constructed by the mixed symmetric tensor $T^{\mu_1 \nu_1 \mu_2 \nu_2}$ and the totally symmetric tensor $N^{\mu_1 \nu_1 \mu_2 \nu_2}$. Then, we can rewrite the action (\ref{nana6}) as follows,
\begin{align}
S=\int d^Dx \sqrt{-g} \left[\frac12 g^{\mu_1 \nu_1 \mu_2 \nu_2 \mu_3 \nu_3}\nabla_{\mu_1} h_{\mu_2 \nu_2 } \nabla_{\nu_1} h_{\mu_3 \nu_3}  
+\frac{1}{2} \left\{ m^2 g^{\mu_1 \nu_1 \mu_2 \nu_2} +T^{\mu_1 \nu_1 \mu_2 \nu_2} +N^{\mu_1 \nu_1 \mu_2 \nu_2} \right\}
h_{\mu_1 \nu_1} h_{\mu_2 \nu_2}\right], \label{nanana7}
\end{align}
with arbitrary mixed symmetric tensor $T^{\mu_1 \nu_1 \mu_2 \nu_2}$ and arbitrary totally symmetric tensor $N^{\mu_1 \nu_1 \mu_2 \nu_2}$.
We perform our calculation with using the symmetries of these tensor $T^{\mu_1 \nu_1 \mu_2 \nu_2}, N^{\mu_1 \nu_1 \mu_2 \nu_2}$, without giving the explicit forms of these tensor, until the explicit forms become necessary.   

\section{CONDITION FOR GHOST-FREENESS}
\label{cond}
\subsection{The condition}
In this part, we perform the Lagrangian analysis in the curved spacetime and derive the condition for the ghost-freeness.
Let us start the Lagrangian analysis in the curved spacetime with arbitrary metric.
The EoM obtained by the variation of the action (\ref{nanana7}) is given by,
\begin{align}
E^{\mu\nu} \equiv \left[-g^{(\mu \nu) \mu_1 \nu_1\mu_2 \nu_2}\nabla_{\mu_2}  \nabla_{\nu_2} 
+ m^2 g^{\mu \nu \mu_1 \nu_1} +T^{\mu \nu \mu_1 \nu_1} +N^{\mu \nu \mu_1 \nu_1} \right]h_{\mu_1 \nu_1} =0. \label{nana7}
\end{align}
Here, we use the symmetries of the tensors $g^{\mu_1 \nu_1 \cdots \mu_n \nu_n},T^{\mu_1\nu_1 \mu_2 \nu_2},N^{\mu_1\nu_1\mu_2 \nu_2}$ given in Eq.(\ref{Appp1}), Eq.(\ref{nana5}) and TABLE \ref{fig1}.
We should note that the symmetrization factor () in the superscript $\mu\nu$ of the kinetic term cannot be eliminated in contrast to the case of the flat spacetime.
Each components of Eq.(\ref{nana7}) are given by,
\begin{align}
&E^{ij} = -g^{(ij) 00 i_1 j_1}\partial_0 \partial_0 h_{i_1j_1} + (\text{terms without }\partial_0 \partial_0 h)^{ij} =0, \label{rr1} \\
&E^{0\nu} = \left[-\frac12 g^{0 \nu i_1 \nu_1 i_2 \nu_2}\nabla_{i_2}  \nabla_{\nu_2} 
-\frac12 g^{\nu 0 \nu_1 i_1\nu_2 i_2}\nabla_{\nu_2}  \nabla_{i_2} 
+ m^2 g^{0 \nu i_1 \nu_1} +T^{0 \nu i_1 \nu_1} \right]h_{i_1 \nu_1}
+N^{0 \nu \mu_1 \nu_1} h_{\mu_1 \nu_1} =0.\label{nana8}
\end{align}
As the same way in the case of the flat spacetime, pure spacial component (\ref{rr1}) can be solved for the acceleration $\partial_0 \partial_0 h_{i_1j_1}$ and the $(0\nu)$ component (\ref{nana8}) can be regarded as the constraint $ E^{0\nu} \equiv \phi^{(1)\nu}  \approx 0$. 
The consistency condition $\dot{\phi}^{(1)\nu} = 0$ can be deformed as the divergence of the EoM $0=\dot{\phi}^{(1)\nu} \approx \nabla_\mu E^{\mu \nu}$.
However, in contrast to the flat space case, the divergence $\nabla_\mu E^{\mu \nu}$ contains some contributions from the kinetic terms due to the non-commutativity of the covariant derivative,
\begin{align}
&\nabla_\mu g^{(\mu \nu) \mu_1 \nu_1\mu_2 \nu_2}\nabla_{\mu_2}  \nabla_{\nu_2} h_{\mu_1 \nu_1} 
=-\left[ \delta T^{\mu\nu\mu_1 \nu_1} + Q^{\mu\nu\mu_1\nu_1} \right] \nabla_\mu
h_{\mu_1 \nu_1} + (\text{terms without any derivatives of }h), \notag \\
&\delta T^{\mu \nu \mu_1 \nu_1 } \equiv - R^{\mu \mu_1 \nu \nu_1} +\left(R^{\mu [\nu} g^{\nu_1] \mu_1 } - R^{\mu_1 [\nu} g^{\nu_1] \mu}\right), \notag \\
&Q^{\mu\nu\mu_1 \nu_1}\equiv \frac12 \left[ R^{\mu\nu} g^{\mu_1 \nu_1} -g^{\mu\nu} R^{\mu_1 \nu_1}
+2 R^{\mu(\mu_1} g^{\nu_1)\nu} -2 R^{\nu(\mu_1}g^{\nu_1) \mu} \right].\label{nana11} 
\end{align}
Here, we decompose the coefficient matrix into the ``integrable term" $\delta T^{\mu\nu\mu_1 \nu_1}$ and the ``non-integrable term" $Q^{\mu\nu\mu_1 \nu_1}$ defined so that those satisfy the following relations, 
\begin{align}
& \frac12 \left[ \delta T^{(\mu\nu)(\mu_1 \nu_1)} + \delta T^{(\mu_1\nu_1)(\mu \nu)}\right] =\delta T^{\mu\nu(\mu_1 \nu_1)},  \notag \\
& \frac12 \left[ Q^{(\mu\nu)(\mu_1 \nu_1)} + Q^{(\mu_1\nu_1)(\mu \nu)}\right] =0.
\end{align}
In other words, the integrable term $\delta T^{\mu\nu\mu_1 \nu_1}$ can be canceled by the contributions from the nonminimal coupling terms and the non-integrable term $Q^{\mu\nu\mu_1 \nu_1}$ cannot be canceled by any other terms.
The existence of the non-integrable terms is the reason why the infinite series of the nonminimal coupling terms are necessary.
By substituting Eq.(\ref{nana11}) into the consistency condition $0=\dot{\phi}^{(2)}\approx \nabla_\mu E^{\mu\nu}$, we obtain,
\begin{align}
&0=\dot{\phi}^{(1)\nu} \approx \nabla_\mu E^{\mu\nu}
= V^{\mu\nu\mu_1 \nu_1} \nabla_\mu h_{\mu_1 \nu_1}
+(\text{terms without any derivatives of }h), 
 \notag \\
&V^{\mu\nu\mu_1 \nu_1} \equiv \left[ m^2 g^{(\mu\nu)(\mu_1 \nu_1)} + \bar{S}^{\mu\nu\mu_1 \nu_1} + N^{\mu \nu \mu_1 \nu_1} + Q^{\mu\nu\mu_1 \nu_1} \right],
 \notag \\
&\bar{S}^{\mu\nu\mu_1 \nu_1} \equiv  S^{(\mu\nu)(\mu_1 \nu_1)}, \notag \\
&S^{\mu\nu\mu_1 \nu_1} \equiv T^{\mu\nu\mu_1 \nu_1} + \delta T^{\mu\nu\mu_1 \nu_1}.
\label{nana13}
\end{align}
We find that the tensor $\delta T^{\mu\nu\mu_1 \nu_1}$ has the mixed symmetry expressed in Eq.(\ref{nana5}).
Thus, we redefine the mixed symmetric tensor $T^{\mu\nu\mu_1 \nu_1}$ as $S^{\mu\nu\mu_1 \nu_1} \equiv T^{\mu\nu\mu_1 \nu_1} + \delta T^{\mu\nu \mu_1 \nu_1}$ in the above equation.
The tensor $S^{\mu_1 \nu_1 \mu_2 \nu_2}$ also has the mixed symmetry such as tensor $T^{\mu_1 \nu_1 \mu_2 \nu_2}$ in Eq.(\ref{nana5}).
Furthermore, we define the ``symmetric bases" $\bar{S}^{\mu\nu\mu_1 \nu_1}$ of the mixed symmetric tensor $S^{\mu\nu\mu_1 \nu_1}$.
From the definition of $\bar{S}^{\mu\nu\mu_1 \nu_1}$ in Eq.(\ref{nana13}) and the symmetry of the anti-symmetric bases of the mixed symmetric tensor given in Eq.(\ref{nana5}), we find that $\bar{S}^{\mu\nu\mu_1 \nu_1}$ satisfies following identities,
\begin{align}
&\bar{S}^{\mu\nu\mu_1 \nu_1} = \frac12 \left[ \bar{S}^{(\mu\nu)(\mu_1 \nu_1)}
+ \bar{S}^{(\mu_1\nu_1)(\mu \nu)}\right], \notag \\
&3\bar{S}^{(\mu_1 \nu_1 \mu_2) \nu_2} =  \bar{S}^{\mu_1 \nu_1 \mu_2 \nu_2}
+\bar{S}^{ \nu_1 \mu_2 \mu_1\nu_2}+\bar{S}^{\mu_2 \mu_1 \nu_1\nu_2}=0. \label{nara1}
\end{align}

Let us back to the Lagrangian analysis.
Eq.(\ref{nana13}) can be regarded as constraint $\phi^{(2)\nu} \equiv \nabla_\mu E^{\mu \nu} \approx 0$. However, we have to restrict the tensors $S^{\mu\nu\mu_1 \nu_1},N^{\mu_1 \nu_1 \mu_2 \nu_2}$ for the existence of the additional constraints.
In order to get the condition, we decompose the first order time derivative terms of Eq.(\ref{nana13}) into $\partial_0 {h}_{ij}$ and $\partial_0 h_{0\mu}$,
\begin{align}
&\nabla_\mu E^{\mu\nu} = \Phi^{\nu \mu} \partial_0 h_{0\mu} + \Psi^{\nu,ij} \partial_0 h_{ij} + (\text{temrs without any time derivatives of }h). \notag \\
&\Phi^{\nu0} \equiv V^{0\nu00},
 \ \ \Phi^{\nu i} \equiv 2V^{0\nu i 0}, \ \ \Psi^{\nu,ij} \equiv V^{0\nu ij}.
\end{align}
We should note that the ``matrix'' $\Phi^{\nu\mu}$ is not a component of the covariant tensor, because there is a difference by factor $2$ between the proportionality factors of the $(\nu0)$-components and the $(\nu i)$-components.  
Let us prove the following proposition.

{\bf Proposition:} If the ``matrix" $\Phi^{\mu\nu}$ is singular, 
\begin{align}
\text{Det} (\Phi^{\mu\nu}g_{\nu\rho}) :=0, \label{yua1}
\end{align}
an additional constraint exists.

{\bf Proof:} The consistency condition of the constraint $\phi^{(2)\nu} \equiv \nabla_\mu E^{\mu\nu}\approx 0$ is given by,
\begin{align}
0= \dot \phi^{(2)\nu} = \partial_0 \nabla_\mu E^{\mu \nu} 
= \Phi^{\nu \mu} \partial_0 \partial_0 h_{0\mu} + \Psi^{\nu,ij} \partial_0 \partial_0 h_{ij} + (\text{temrs without any second time derivatives of }h). \label{yua2}
\end{align}
By the assumption (\ref{yua1}), the matrix $\Phi^{\mu\nu}$ has a zero eigenvector $u_\mu$ satisfying the relation $u_{\mu }\Phi^{\mu\nu} =0$. 
Operating the zero eigenvector $u_{\nu}$ on Eq.(\ref{yua2}), 
we obtain the equation without any second order time derivatives of $h_{0\mu}$,
\begin{align}
0= u_{\nu} \dot \phi^{(2)\nu} 
= u_\nu \Psi^{\nu,ij} \partial_0 \partial_0 h_{ij} + (\text{temrs without any second time derivatives of }h). 
\end{align}
The second order time derivatives of $h_{ij}$ can be eliminated by Eq.(\ref{nana8}).
Then, we obtain an additional constraint if the condition (\ref{yua1}) is satisfied.

\subsection{Cofactor expansion}
In the later section, we will try to solve the condition (\ref{yua1}) by considering the perturbation with respect to $R/m^2$. However, the straightforward expansion of the condition (\ref{yua1}) force tedious calculations on us.
In this part, we deform the condition (\ref{yua1}).

First, we can easily show the following relation,
\begin{align}
\text{Det} (\Phi^{\mu\nu}g_{\nu\rho}) = 2^{D-1}\text{Det}(V^{0\mu\nu0}g_{\nu\rho}).
\end{align}
Then we can rewrite the condition (\ref{yua1}) as follows,
\begin{align}
\text{Det}(V^{0\mu\nu0}g_{\nu\rho}):=0,\label{yua3}
\end{align}
by using the covariant tensor $V^{0\mu\nu0}$.

Next, by noting the relation $g^{(0 \mu)(\nu 0)} = \frac12 g^{0\mu \nu0}$,
$V^{0\mu\nu0}$ can be expressed as follows,
\begin{align}
&V^{0\mu\nu0}= -\frac{m^2}{2} g^{00} \theta^{\mu\nu} +  \bar{S}^{0\mu \nu 0} + N^{0\mu \nu 0}
+Q^{0\mu \nu 0}, \notag \\
&\theta^\mu_\nu \equiv \delta^\mu_\nu - \frac{g^{\mu0} g^0_\nu}{g^{00}},
 \ \ Q^{0\mu\nu0}= \frac12 \left[ R^{00}g^{\mu\nu} - g^{00}R^{\mu\nu} \right]. \label{yua4}
\end{align}
Here, the matrix $\theta^\mu_\nu$ is the projection operator living in the $D-1$ dimensional space orthogonal to $g^0_\mu$.
We lower the indexes of $\theta^\mu_\nu$ by using the metric $g_{\mu\nu}$ and raise the indexes by using the inverse matrix $g^{\mu\nu}$.
Furthermore, we should note that the tensor $V^{0\mu\nu0}$ is symmetric with respect to the superscripts $\mu\nu$.

We would like to consider the $1+(D-1)$ decomposition of the condition (\ref{yua3}).
By decomposing the tensor $V^{0\mu\nu0} $ in Det$(V^{0\mu\nu0} g_{\nu\rho})$ as $V^{0\mu\nu0}= \left( \theta^\mu_\alpha + \frac{g^{0\mu}g^0_\alpha}{g^{00}} \right)V^{0\alpha \beta0} \left(\theta_\beta^\nu + \frac{g^{0\nu}g^0_\beta}{g^{00}}\right) $, we can show the following relation\footnote{We use the relation (\ref{klkl1}) given in Appendix \ref{ap1}},
\begin{align}
\text{Det}(V^{0\mu \nu0}g_{\nu \rho}) =& \frac{1}{D!} g^{\mu_1 \nu_1 \mu_2 \nu_2 \cdots \mu_D \nu_D}V^{0~~~~~0}_{~~\mu_1\nu_1}V^{0~~~~~0}_{~~\mu_2\nu_2} \cdots
 V^{0~~~~~~~0}_{~~\mu_D\nu_D} \notag \\
=&V^{0000} \text{Det}_\theta (V^{0\mu \nu0}\theta_{\nu \rho})  - V^{0~~00}_{~~\mu} V^{00~~0}_{~~~\nu} Y^{\mu\nu}_\theta (V^{0\alpha \beta0}).
\end{align}
Here, we define the $D-1$ dimensional determinant Det${}_\theta (V^{0\mu \nu0}\theta_{\nu \rho})$ and the $D-1$ dimensional cofactor matrix $Y^{\mu\nu}_\theta (V^{0\alpha \beta0})$ as follows,
\begin{align}
&\text{Det}_\theta (V^{0\mu \nu0}\theta_{\nu \rho}) \equiv \frac{1}{(D-1)!} \theta^{\mu_2 \nu_2 \cdots \mu_D\nu_D} V^{0~~~~~0}_{~~\mu_2\nu_2} \cdots V^{0~~~~~~~0}_{~~\mu_D\nu_D}, \notag \\
&Y^{\mu\nu}_\theta (V^{0\alpha \beta0}) \equiv \frac{1}{(D-2)!}  \theta^{\mu \nu \mu_3 \nu_3 \cdots \mu_D \nu_D} 
 V^{0~~~~~0}_{~~\mu_3\nu_3} \cdots V^{0~~~~~~~0}_{~~\mu_D\nu_D}, \notag \\
&\theta^{\mu_1 \nu_1 \mu_2 \nu_2 \cdots \mu_n \nu_n}
\equiv \delta^{\mu_1~~\mu_2~~ \cdots \mu_n }_{~~~\rho_1~~\rho_2 ~~\cdots \rho_n}
\theta^{\rho_1\nu_1} \theta^{\rho_2 \nu_2} \cdots \theta^{\rho_n \nu_n}.
\end{align}
Perturbatively, the $D-1$ dimensional determinant is not equal to zero Det${}_\theta (V^{0\mu\nu0} \theta_{\nu \rho}) \neq 0$.
Then we obtain the condition,
\begin{align}
&V^{0000} - V^{0~~00}_{~~\mu} V^{00~~0}_{~~~\nu} (\theta V \theta)^{-1 \nu \mu} :=0,\notag \\
&(\theta V \theta)^{-1 \nu\mu} \equiv \frac{Y^{\mu\nu}_\theta (V^{0\alpha \beta0}) }{\text{Det}_\theta (V^{0\mu \nu0}\theta_{\nu \rho}) }. \label{yui1}
\end{align}
We can easily show that the matrix $(\theta V \theta)^{-1 \mu\nu}$ defined in the above equation is the ``inverse matrix" satisfying the following identity
\footnote{The identity (\ref{yua8}) can be shown by the trivial identity (Hamilton-Cayley's theorem) given as follows,
\begin{align}
0&= \theta^{\mu_1~~\mu_2 ~~\cdots \mu_D }_{~~~\nu_1~~\nu_2~~\cdots \nu_D} \notag \\
&= D\theta^{\mu_1 }_{~[\nu_1} \theta^{\mu_2~~\cdots \mu_D }_{~~~\nu_2~~\cdots \nu_D]}. \label{yua7}
\end{align}
This identity is due to the dimension of the projection operator $\theta^{\mu\nu}$.
In the second line, we expand $\theta^{\mu_1~~\mu_2 ~~\cdots \mu_D }_{~~~\nu_1~~\nu_2~~\cdots \nu_D}$ such as expansion given in (\ref{Ap3}) in the Appendix \ref{ap1}.
By operating the product $ V^{0~~~\nu_2 0}_{~~\mu_2} \cdots  V^{0~~~\nu_D0}_{~~\mu_D}$ 
on the identity (\ref{yua7}), we obtain the identity (\ref{yua8}).},
\begin{align}
(\theta V \theta)^{-1\mu\nu} \theta_{\nu \alpha} V^{0\alpha \beta0} \theta_{\beta \rho}
= \theta^\mu_\rho. \label{yua8}
\end{align}

Let us deform the ``inverse matrix" satisfying the above identity.
We would like to decompose the matrix $V^{0\mu\nu0}$ into the mass terms and the other terms $\Delta^{\mu\nu}$ as follows,
\begin{align}
&V^{0\mu\nu0}= -\frac{m^2}{2} g^{00} \theta^{\mu\nu} +  \Delta^{\mu\nu}, \notag \\
&\Delta^{\mu\nu} \equiv  \bar{S}^{0\mu \nu 0} + N^{0\mu \nu 0}
+Q^{0\mu \nu 0}. \label{saya1}
\end{align}
Substituting the above expression into the identity (\ref{yua8}), we obtain,
\begin{align}
(\theta V \theta)^{-1\mu\nu} \left( -\frac{m^2}{2} g^{00} \theta_{\nu\rho} +  \theta_{\nu \alpha} \Delta^{\alpha\beta} \theta_{\beta\rho} \right)
= \theta^\mu_\rho.
\end{align}
By using the constraints $(\theta V \theta)^{\mu0}=0=(\theta V \theta)^{0\mu}$ obtained from the definition (\ref{yui1}), we can uniquely solve the above equation as follows,
\begin{align}
(\theta V \theta)^{-1 \mu\nu} = -
 \sum_{n=0}^\infty \left( \frac{2}{m^2g^{00}} \right)^{n+1} \left[\theta(\Delta \theta)^n\right]^{\mu\nu}. \label{saya2}
\end{align}
Of course, the expression (\ref{saya2}) and the definition (\ref{yui1}) are related with each other through Hamilton-Caley's theorem (\ref{yua7}) given in the footnote. 
Substituting (\ref{saya1}) and (\ref{saya2}) into the condition (\ref{yui1}), 
we obtain the following condition,
\begin{align}
\Delta^{00} +  
 \sum_{n=0}^\infty \left( \frac{2}{m^2g^{00}} \right)^{n+1} \Delta^0_{~\nu} \left[\theta( \Delta \theta)^n\right]^{\nu\mu}\Delta_\mu^{~0}   :=0. \label{saya3}
\end{align}

\section{PERTURBATIVE SOLUTION}
\label{pert}
In this section, we derive the perturbative solution of the condition (\ref{saya3}).

Let us expand the tensors $\bar{S}^{\mu_1 \nu_1 \mu_2 \nu_2},N^{\mu_1 \nu_1 \mu_2 \nu_2}$ in powers of $R/m^2$ as follows,
\begin{align}
&\bar{S}^{\mu_1 \nu_1 \mu_2 \nu_2}= \sum_{n=1}^\infty \frac{1}{m^{2(n-1)}}\bar{S}^{(n)\mu_1 \nu_1 \mu_2  \nu_2}, \notag \\
&N^{\mu_1 \nu_1 \mu_2 \nu_2}= \sum_{n=1}^\infty \frac{1}{m^{2(n-1)}}N^{(n)\mu_1 \nu_1 \mu_2  \nu_2}. \label{kana1}
\end{align}
Here, the superscript $(n)$ means the $n$ th order term with respect to curvatures.

If we substitute the expression (\ref{kana1}) into the condition (\ref{saya3}),
we can regard the condition (\ref{saya3}) as the condition which determine the totally symmetric tensor $N^{(n)\mu_1 \nu_1 \mu_2 \nu_2}$ in order by order.
Let us explain this fact.

Eq.(\ref{saya3}) is still invariant under exchanging the totally symmetric tensor $N^{\mu_1 \nu_1 \mu_2 \nu_2}$ and the mixed symmetric tensor $\bar{S}^{\mu_1 \nu_1 \mu_2 \nu_2}$ because these tensors appear only as the combination $N^{\mu_1 \nu_1 \mu_2 \nu_2} + \bar{S}^{\mu_1 \nu_1 \mu_2 \nu_2}$.
However, this invariance is broken by using the the mixed symmetry of $\bar{S}^{\mu_1 \nu_1 \mu_2 \nu_2}$.
From the identity Eq.(\ref{nara1}), we find that the mixed symmetric tensor $\bar{S}^{\mu_1 \nu_1 \mu_2 \nu_2}$
 cannot has same superscripts whose number is more than two, i.e.,
\begin{align}
\bar{S}^{00\mu0}=0=\bar{S}^{0\mu00}. \label{saya5}
\end{align}
Then, we find that the mixed symmetric tensor $\bar{S}^{0\mu\nu0}$ depends only on the pure spatial components for the $1+(D-1)$ decomposition,
\begin{align}
\bar{S}^{0\mu\nu0} = \theta^\mu_\alpha \bar{S}^{0\alpha \beta 0} \theta_\beta^{\nu}.
\label{yua5}
\end{align}
Substituting the identity (\ref{saya5}) or (\ref{yua5}) into the condition (\ref{saya3}),
the invariance under the exchange $N^{\mu_1 \nu_1 \mu_2 \nu_2} \leftrightarrow \bar{S}^{\mu_1 \nu_1 \mu_2 \nu_2}$ is broken,
\begin{align}
\hat{N}^{00} +  
 \sum_{n=0}^\infty \left( \frac{2}{m^2g^{00}} \right)^{n+1} \left(\hat{N} +\hat{Q} \right)^0_{~\nu} \left[\theta( \Delta \theta)^n\right]^{\nu\mu}\left(\hat{N}+\hat{Q}\right)_\mu^{~0}   :=0.
\label{nara2}
\end{align}
Here, we defined the matrix $\hat{N}^{\mu\nu} \equiv N^{0\mu\nu0}, \ \hat{S}^{\mu\nu} \equiv \bar{S}^{0\mu\nu0}, \ \hat{Q} ^{\mu\nu}\equiv Q^{0\mu\nu0}$ and the product $[XY]^{\mu\nu} \equiv X^{\mu\rho} Y_{\rho}^{~\nu}$.  
We find that Eq.(\ref{nara2}) can be solved for $N^{(n)0000}$, in order by order, as some functions of the lower order quantity $N^{(m)0 \mu \nu 0 }, S^{(m)0\mu\nu0} (m\le n)$ and the tensor $Q^{0\mu\nu0}$.
Thus, in principle, we can regard the condition (\ref{nara2}) as a constraint on the totally symmetric tensor.

\subsection{Lemma}
\label{lem}
Before beginning our calculation, we would like to give a lemma which seems a little bit trivial but plays a crucial role in later analysis. 
All the resulting forms of the conditions in order by order take the following expression,
\begin{align}
D^{0000}:=0. \label{shi1}
\end{align}
Here, $D^{\mu_1 \nu_1 \mu_2 \nu_2}$ is a fourth rank covariant tensor.
Eq.(\ref{shi1}) can be regarded as the condition which constrain not only the $0000$-component $D^{0000}$ but also its totally symmetric parts $D^{(\mu_1 \nu_1 \mu_2 \nu_2)}$,
because we can prove the following lemma.

{\bf Proposition:} We assume the tensor $D^{\mu_1 \nu_1 \mu_2 \nu_2}$ satisfying the following properties:
\begin{enumerate}
\item
$D^{\mu_1 \nu_1 \mu_2 \nu_2}$ depends only on the metric $g_{\mu\nu}$ and its partial  derivatives.
\item
$D^{\mu_1 \nu_1 \mu_2 \nu_2}$ is general covariant.
\item
For ``any metric", the 0000-component of the tensor $D^{\mu_1 \nu_1 \mu_2 \nu_2}$ is equal to zero,
\begin{align}
D^{0000}(g_{\mu\nu}):=0. \label{lem1}
\end{align}

\end{enumerate}
Under the above assumption, the totally symmetric parts $D^{(\mu_1 \nu_1 \mu_2 \nu_2)}$ must be equal to zero.

{\bf Proof:}
The assumption 3 means that the variation of Eq.(\ref{lem1}) with respect to the metric is also equal to zero,
\begin{align}
 D^{0000}(g_{\mu\nu} + \delta g_{\mu\nu})
- D^{0000}(g_{\mu\nu}): =0.\label{lem2}
\end{align}
because we assume Eq.(\ref{lem1}) is valid for ``any metric".
We choose the variation $\delta g_{\mu\nu}$ as the Lie-derivative of the metric, i.e., $\delta g_{\mu\nu} = 2\nabla_{(\mu} \xi_{\nu)}$.
Then, using the assumption 1,
Eq.(\ref{lem2}) can be rewritten in terms of the Lie-derivative of $D^{0000}$,\footnote{We should note that the Lie-derivative commutate with the partial derivative 
$[\mathcal{L}_\xi,\partial_\mu]=0$, in contrast to the infinitesimal general coordinate transformation. Thus, this statement is valid even if $D^{0000}$ contains the curvatures and/or its covariant derivatives.}
\begin{align}
\mathcal{L}_{\xi}{D}^{0000}: = 0.
\end{align}
On the other hand, from the assumption 2, 
the Lie-derivative $\mathcal{L}_\xi D^{0000}$ is given by,
\begin{align}
\mathcal{L}_\xi D^{0000}&= \xi^\alpha \nabla_\alpha D^{0000} - 4D^{(\rho 000)} \nabla_\rho \xi^0\notag \\
&=\xi^\alpha \partial_\alpha D^{0000} - 4D^{(\rho 000)} \partial_\rho \xi^0.
\end{align}
$\xi^\alpha \partial_\alpha D^{0000}$ in the above equation is equal to zero under Eq.(\ref{lem1}).
Then, the second term must be equal to zero for any $\xi^0$.
Hence, we obtain,
\begin{align}
D^{(\rho 000)} :=0.
\end{align}
As the same way, the additional Lie-derivative of the above equation gives,
\begin{align}
D^{(\rho \sigma 00)}:=0.
\end{align}
By repeating the above procedure, we obtain,
\begin{align}
D^{(\mu_1 \nu_1 \mu_2 \nu_2)}:=0.
\end{align}
Therefore the proposition is valid.

\subsection{Leading order}
Let us investigate the perturbative conditions.
By substituting the expression (\ref{kana1}) into Eq.(\ref{nara2}),
and comparing the both sides of Eq.(\ref{nara2}),
we obtain the following condition in the leading order,
\begin{align}
N^{(1)0000}:=0. \label{kana2}
\end{align}
From the lemma given in the previous part, Sec.\ref{lem}, we find that the full components of the totally symmetric tensor $N^{(1)\mu_1 \nu_1 \mu_2 \nu_2}$ must be equal to zero,
\begin{align}
N^{(1)\mu_1 \nu_1 \mu_2 \nu_2} :=0. \label{kana3}
\end{align}
Simultaneously, we find that the mixed symmetric tensor $S^{(1)\mu_1 \nu_1 \mu_2 \nu_2}$
\footnote{As the same way in the case of $\bar{S}^{\mu_1 \nu_1 \mu_2 \nu_2},N^{\mu_1 \nu_1 \mu_2 \nu_2}$, we define the expansion of the tensors $S^{\mu_1 \nu_1 \mu_2 \nu_2}, T^{\mu_1 \nu_1 \mu_2 \nu_2}$ as,
\begin{align}
S^{\mu_1 \nu_1 \mu_2 \nu_2}= \sum_{n=1}^\infty \frac{1}{m^{2(n-1)}}S^{(n)\mu_1 \nu_1 \mu_2  \nu_2},  \ \ \ T^{\mu_1 \nu_1 \mu_2 \nu_2}= \sum_{n=1}^\infty \frac{1}{m^{2(n-1)}}T^{(n)\mu_1 \nu_1 \mu_2  \nu_2}. 
\end{align}
}
is not constrained by the leading order condition (\ref{kana2}), because the mixed symmetric tensor vanish from the leading order condition (\ref{kana2}) due to its symmetry (\ref{saya5}). Then, the mixed symmetric tensor $T^{(1)\mu_1 \nu_1 \mu_2 \nu_2}$ in Eq.(\ref{nanana7}) is not also constrained by the leading order condition, because the both tensors $T^{\mu_1 \nu_1 \mu_2 \nu_2},S^{\mu_1 \nu_1 \mu_2 \nu_2}$ are related with each other through 
Eq.(\ref{nana11}).
This result perfectly coincide with Buchbinder's result in Eq.(\ref{int1}), because the nonminimal coupling terms in the action (\ref{int1}) are the most general representation of the mixed symmetric tensor $T^{\mu_1 \nu_1 \mu_2 \nu_2}$ in the leading order.

\subsection{Second order}
Pure original results in this paper are the constraints from the second or higher order conditions.
In the second order, the straightforward expansion of the condition (\ref{nara2}) is given by,
\begin{align}
N^{(2)0000} + \frac{2}{g^{00}} Q^{00\mu0} \theta_{\mu\nu} Q^{0\nu00}:=0.\label{dd1}
\end{align}
Here, we use the results of the leading order condition (\ref{kana3}). 
From Eq.(\ref{yua4}), the explicit form of the tensor $Q^{00\mu0}=Q^{0\mu00}$
is given by,
\begin{align}
Q^{00\mu0} = -\frac12 g^{00} \left[\theta R \right]^{\mu0}.\label{dd2}
\end{align}
\if0
By substituting Eq.(\ref{dd2}) into the condition (\ref{dd1}), we obtain,
\begin{align}   
N^{(2)0000} + \frac{g^{00}}{2}  R^{0\mu} \theta_{\mu\nu} R^{\nu0}:=0.
\end{align}
\fi
By substituting Eq.(\ref{dd2}) into the condition (\ref{dd1}) and using the definition of $\theta^{\mu\nu}$ in Eq.(\ref{yua4}),
we obtain the second order condition,
\begin{align}
N^{(2)0000}+\frac12 g^{00\mu\nu}R^0_\mu R^0_\nu :=0.\label{nara4}
\end{align}
From the lemma give in Sec.\ref{lem}, 
the full components of $N^{(2)\mu_1 \nu_1 \mu_2 \nu_2}$ are uniquely determined as follows,
\begin{align}
N^{(2)\mu_1 \nu_1 \mu_2 \nu_2} := -\frac12 g^{\alpha \beta (\mu_1 \nu_1}R^{\mu_2}_\alpha R^{\nu_2)}_\beta. \label{kana4}
\end{align}

We find that the nonminimal coupling terms cannot be truncated at the leading order, because $N^{(2)\mu_1 \nu_1 \mu_2 \nu_2}$ must be not equal to zero no mater how we chose the mixed symmetric tensors $S^{(n)\mu_1 \nu_1 \mu_2 \nu_2}$.
Then, we have confirmed the necessity of the nonminimal coupling terms higher than the leading order.

Similarly to the case of the leading order, the second order condition does not constrain any mixed symmetric tensors $S^{(n)\mu_1 \nu_1 \mu_2 \nu_2}$.

\subsection{Third order}
In the third order, the straightforward expansion is given by,
\begin{align}
N^{(3)0000} + \frac{4}{g^{00}} \left[\hat{N}^{(2)}\theta \hat{Q}\right]^{00} 
+\left(\frac{2}{g^{00}} \right)^2 \left[ \hat{Q} \left( \hat{S}^{(1)} + \hat{Q} \right) \hat{Q}\right]^{00} :=0.
\label{dd3}
\end{align}
Here, we define the matrices $\bar{S}^{(n)0\mu\nu0}\equiv \hat{S}^{(n)\mu\nu}, \ \bar{N}^{(n)0\mu\nu0} \equiv \hat{N}^{(n)\mu\nu}$.
Furthermore, we omit the projection operator $\theta$ from the third term in the left hand side of the above equation because the inner products of the third term is automatically projected due to Eq.(\ref{dd2}).
From the second order result Eq.(\ref{kana4}) and the explicit form of $Q^{0\mu\nu0}$ in Eq.(\ref{yua4}), we can easily show the following identity,
\begin{align}
\left[ N^{(2)00\mu0} + \frac{1}{g^{00}} Q^{0\mu\nu0}Q^{0~~00}_{~~\nu} \right] 
\theta_\mu^{\sigma} =0. \label{dd5}
\end{align}
Using the above identity, the condition (\ref{dd3}) can be deformed as follows,
\begin{align}
N^{(3)0000} + \left(\frac{2}{g^{00}} \right)^2 \left[ \hat{Q} \hat{S}^{(1)} \hat{Q}\right]^{00}:=0.\label{lie5}
\end{align}
By substituting Eq.(\ref{dd2}) into the above expression and using the identity (\ref{yua5}),
we obtain the third order condition,
\begin{align}
N^{(3)0000} +\bar{S}^{(1)0\alpha \beta0} R^0_\alpha R^0_\beta :=0.
\end{align}
Then, the full components are uniquely determined as follows,
\begin{align}
N^{(3)\mu_1 \nu_1 \mu_2 \nu_2} :=  \frac12 S^{(1)\alpha \beta (\mu_1 \nu_1} 
R^{\mu_2}_\alpha R^{\nu_2)}_\beta. \label{nara3}
\end{align}
Here, we rewrite the above equation in terms of the anti-symmetric bases $S^{(1)\mu_1 \nu_1 \mu_2 \nu_2}$ by using the relation $ \bar{S}^{(1)\mu \alpha \beta \nu} + \bar{S}^{(1)\nu \alpha \beta \mu}  = -\bar{S}^{(1)\alpha \beta \mu\nu}= - S^{(1)(\alpha \beta) (\mu\nu)}$, which is obtained from the symmetries (\ref{nara1}) and the definition (\ref{nana13}).

In contrast to the case of the lower order,
the third order totally symmetric tensor $N^{(3)\mu_1 \nu_1 \mu_2 \nu_2}$ depends on the leading order mixed symmetric tensor $S^{(1)\mu_1 \nu_1 \mu_2 \nu_2}$.
Then, we find that the totally symmetric tensors $N^{(n)\mu_1 \nu_1 \mu_2 \nu_2}$ are generally some functions of the mixed symmetric tensors $S^{(n)\mu_1 \nu_1 \mu_2 \nu_2}$.

The mixed symmetric tensors are not also constrained in the third order.

\subsection{Identities from Lie-derivative}
\label{lieder}
In the previous part, we obtained the identity (\ref{dd5}) straightforwardly by using the second order result (\ref{kana4}).
We can derive the identity (\ref{dd5}) more systematically by taking the Lie-derivative of Eq.(\ref{dd1}).

From Eq.(\ref{dd2}), we find the identity $Q^{00\mu0}=\theta^\mu_\nu Q^{00\nu0}$.
Using this identity, Eq.(\ref{dd1}) can be rewritten as follows,
\begin{align}
N^{(2)0000} + \frac{2}{g^{00}} Q^{00\mu0} Q^{0~~00}_{~~\mu}:=0.\label{lie2}
\end{align}
Taking the Lie-derivative of the above equation,
we obtain,
\begin{align}
0= \xi^\alpha \partial_\alpha \left[ N^{(2)0000} + \frac{2}{g^{00}} Q^{00\mu0} Q^{0~~00}_{~~\mu} \right] 
- \left[ 4 N^{(2)000\rho}  + \frac{2}{g^{00}}\cdot 6 Q^{00\mu0} Q^{(\rho~~00)}_{~~\mu} 
- \frac{4}{(g^{00})^2} g^{\rho 0} Q^{00\mu0} Q^{0~~00}_{~~\mu} 
\right]
\partial_\rho \xi^0.\label{lie1}
\end{align}
Here, we use the totally symmetric property of $N^{(2)\mu_1 \nu_1 \mu_2 \nu_2}$. The first term of the right hand side in Eq.(\ref{lie1}) becomes equal to zero under the original condition (\ref{lie2}).
Thus, the second term must be equal to zero ``for any $\xi^0$".
So, we obtain,
\begin{align}
0= 4 N^{(2)000\rho}  + \frac{2}{g^{00}}\cdot 6 Q^{00\mu0} Q^{(\rho~~00)}_{~~\mu} 
- \frac{4}{(g^{00})^2} g^{\rho 0} Q^{00\mu0} Q^{0~~00}_{~~\mu}. 
\end{align}
Operating the projection operator $\theta$ on the above equation and using the identity $Q^{00\mu0}=Q^{00\nu0}\theta_\nu^\mu$,
we obtain,
\begin{align}
0= \left[ N^{(2)000\rho}  
+ \frac{1}{g^{00}}\cdot 3 Q^{00\nu0} \theta_\nu^\mu
 Q^{(\rho~~00)}_{~~\mu} \right] \theta_\rho^\sigma.  \label{lie3}
\end{align}
$Q^{(\rho~~00)}_{~~\mu}$ is given by,
\begin{align}
3Q^{(\rho~~00)}_{~~\mu}= R^{0\rho} g^0_\mu - R^{0}_\mu g^{0\rho} + Q^{0~~\rho0}_{~\mu}.
\end{align}
Operating the projection operator $\theta$ on the above equation, we obtain,
\begin{align}
3\theta^\alpha_\rho \theta^{\beta\mu} Q^{(\rho~~00)}_{~~\mu}
= \theta^\alpha_\mu Q^{0\mu\nu0} \theta_\nu^\beta.\label{lie4}
\end{align}
Substituting Eq.(\ref{lie4}) into Eq.(\ref{lie3}), we obtain the relation,
\begin{align}
0= \left[ N^{(2)000\rho}  
+ \frac{1}{g^{00}}\cdot  Q^{00\mu0} 
 Q^{0~~\rho0}_{~~\mu} \right] \theta_\rho^\sigma.
\end{align}
This relation can be regarded as an ``identity" under the second order results (\ref{kana4}) because (\ref{kana4}) is derived by using the lemma given in Sec.\ref{lem}.

As the same way, we obtain the identity corresponding to the condition (\ref{lie5}),
\begin{align}
\left[ N^{(3)000\rho} + \frac{2}{(g^{00})^2} \left\{ Q^{0\rho\mu0}
 \bar{S}^{(1)0~~~~0}_{~~~~~\mu\nu} Q^{0\nu00} + Q^{00\mu0}  \bar{S}^{(1)(\rho~~~~0)}_{~~~~~~\mu\nu} Q^{0\nu00} \right\}\right] \theta_{\rho\sigma} =0.\label{lie6}
\end{align}

\subsection{Fourth order}
In the fourth order, the straightforward expansion is given by,
\if0
\begin{align}
-N^{(4)0000} =& \frac{2}{g^{00}} \left\{ 
2\left[ N^{(3)} \theta Q \right]^{00} + \left[ N^{(2)} \theta N^{(2)} \right]^{00}
\right\}
+ \left(\frac{2}{g^{00}}\right)^2\left\{ 2\left[N^{(2)} \theta \left( \bar{S}^{(1)} + Q \right) Q \right]^{00}
+ \left[Q\left( N^{(2)} + \bar{S}^{(2)} \right) Q \right]^{00}
\right\} \notag \\
&+ \left(\frac{2}{g^{00}}\right)^3 \left[ Q\left( \bar{S}^{(1)} + Q\right) \theta\left ( \bar{S}^{(1)} + Q\right) Q\right]^{00}.
\end{align}
\fi
\begin{align}
&0:= N^{(4)0000} + \mathcal{A} + \mathcal{B} +\left( \frac{2}{g^{00}} \right)^2[\hat{Q}\hat{S}^{(2)}\hat{Q}]^{00} 
+ \left( \frac{2}{g^{00}} \right)^3 \left[ \hat{Q}\left(\hat{S}^{(1)} \right)^2 \hat{Q}\right]^{00}, 
\notag \\
&\mathcal{A}\equiv \frac{2}{g^{00}} \left[ \hat{N}^{(2)} \theta \hat{N}^{(2)} \right]^{00}
+ \left(\frac{2}{g^{00}}\right)^2\left\{ 2\left[\hat{N}^{(2)} \theta  \hat{Q}^2 \right]^{00}
+ \left[\hat{Q} \hat{N}^{(2)} \hat{Q} \right]^{00}\right\} 
+\left(\frac{2}{g^{00}}\right)^3 \left[ \hat{Q}^2 \theta \hat{Q}^2 \right]^{00}, \notag \\
&\mathcal{B} \equiv \frac{4}{g^{00}} 
\left[ \hat{N}^{(3)}  \hat{Q} \right]^{00} 
+ 2\left(\frac{2}{g^{00}}\right)^2 \left[\hat{N}^{(2)} \hat{S}^{(1)}  \hat{Q} \right]^{00}
+ 2 \left(\frac{2}{g^{00}}\right)^3 \left[ \hat{Q} \hat{S}^{(1)} \hat{Q}^2 \right]^{00}.\label{kyo3}
\end{align}
Here, we define $\mathcal{A}$ and $\mathcal{B}$ as the terms independently of $\bar{S}^{(n)\mu_1 \nu_1 \mu_2 \nu_2}$, and the terms linear in $\bar{S}^{(1)}$, respectively. 

Using the identity (\ref{dd5}), the quantity $\mathcal{A}$ can be simplified as follows,
\begin{align}
\mathcal{A} &= \frac{2}{g^{00}} \left[ \hat{N}^{(2)} \theta \hat{N}^{(2)} \right]^{00}
+ \left(\frac{2}{g^{00}}\right)^2  \left[\hat{Q} \hat{N}^{(2)}\hat{Q} \right]^{00}.\notag \\
&= \frac18 g^{00\alpha \beta} \left(R^2\right)^0_\alpha \left( R^2\right)^0_\beta
+ N^{(2)0\alpha \beta0} R^0_\alpha R^0_\beta. \label{kyo1}
\end{align}
Here, we substitute the explicit form of $N^{00\mu0}$ and $Q^{00\mu0}$ for the deformation between the first line and second line.

Next, let us calculate $\mathcal{B}$.
By operating $Q^{0\sigma00}$ on the identity (\ref{lie6}) and using the identity $\bar{S}^{(\mu_1 \nu_1 \mu_2) \nu_2} =0$, we obtain the following identity,
\begin{align}
\left[ \hat{N}^{(3)} \hat{Q} \right]^{00} + \frac{2}{(g^{00})^2}\left[ \hat{Q}^2 \hat{S}^{(1)} \hat{Q} \right]^{00} =0.\label{fot1}
\end{align}
By using the above identity (\ref{fot1}) and the identity (\ref{dd5}),
we can easily show that the quantity $\mathcal{B}$ becomes equal to zero,
\begin{align}
\mathcal{B}=0.\label{kyo2}
\end{align}
Finally, substituting the identities (\ref{kyo1}) and (\ref{kyo2}) into the condition (\ref{kyo3}), we obtain,\footnote{Although the explicit form of $N^{(2) 0\alpha \beta 0} R^0_\alpha R^0_\beta$ is given by,
\begin{align}
N^{(2) 0\alpha \beta 0} R^0_\alpha R^0_\beta
&= -\frac{1}{12} \left[ 2g^{\rho \sigma \alpha 0} R^0_\alpha R^0_\sigma 
\left(R^2 \right)^0_\rho +g^{\rho \sigma 00}
 \left(R^2 \right)^0_\rho\left(R^2 \right)^0_\sigma \right] \notag \\
&= -\frac{1}{12} \left[ g^{00} (R^4)^{00} +2R^{00} (R^3)^{00} -3 (R^2)^{00}(R^2)^{00} \right],
\end{align}
this explicit form is not so clusial for our argument.
Then, we do not substitute this explicit form.}
\begin{align}
&N^{(4)0000}+\left( \bar{S}^{(2)0\alpha \beta 0}+N^{(2) 0\alpha \beta 0} \right)R^0_\alpha R^0_\beta
 \notag \\
&+\frac18 g^{00\alpha \beta} \left(R^2\right)^0_\alpha \left(R^2\right)^0_\beta
 + \frac{2}{g^{00}} R^{0\nu} \bar{S}^{(1)0~~\rho0}_{~~~~~\nu}\bar{S}^{(1)0~\sigma 0}_{~~~~\rho} R^0_\sigma:=0.\label{fu6}
\end{align}

We find an obvious difference between the forth order condition and the lower order conditions.
The left hand side of the condition (\ref{fu6}) contains a non-covariant term, which is the forth term.
In general, this non-covariant term cannot be cancelled by the other terms.
This fact implies that the mixed symmetric tensor $S^{(1)\mu_1 \nu_1 \mu_2 \nu_2}$ should be constrained so that the numerator of the non-covariant term satisfies the following condition,
\begin{align}
 2R^{0\nu} \bar{S}^{(1)0~~\rho0}_{~~~~~\nu}\bar{S}^{(1)0~\sigma 0}_{~~~~\rho} R^0_\sigma := 
g^{00} M^{(4)0000}, \label{sr1}
\end{align}
for some covariant tensor $M^{(4)\mu_1 \nu_1 \mu_2 \nu_2}$.
Although the mixed symmetric tensor has the three bases generally as given in Eq,(\ref{nana4}),
the condition (\ref{sr1}) reduces the three bases to the following two bases,
\begin{align}
S^{(1)\mu_1 \nu_1 \mu_2 \nu_2} :=\gamma^{(1)}_1 R g^{\mu_1 \nu_1 \mu_2 \nu_2} 
+\frac{\gamma^{(1)}_2}{2} \left( R^{\mu_1 [\nu_1} g^{\nu_2] \mu_2 } - R^{\mu_2 [\nu_1} g^{\nu_2] \mu_1}\right). \label{yui2}
\end{align}
Here, $\gamma^{(1)}_1, \gamma^{(1)}_2$ are the free parameters.
\footnote{Under the condition (\ref{yui2}), $M^{(4)0000}$ defined in (\ref{sr1}) is given by,
\begin{align}
M^{(4)0000} = \frac{\left(\gamma_1^{(1)}\right)^2}{2^2}
 R^2 g^{00\alpha \beta} R^0_\alpha R^0_\beta
-\frac{\gamma_1^{(1)}\gamma_2^{(1)}}{2^3} R g^{00\alpha \beta} R^0_\alpha (R^2)_\beta^0
+\frac{\left(\gamma_2^{(1)}\right)^2}{2^6}
 g^{00\alpha \beta} (R^2)^0_\alpha (R^2)^0_\beta.
\end{align}}
Therefore, the possibility of the Reimann curvature $R^{\mu_1 \nu_1 \mu_2 \nu_2}$ is excluded.
Substituting the expression (\ref{sr1}) into the condition (\ref{fu6}),
we find that the full components of the totally symmetric tensor $N^{(4)\mu_1 \nu_1 \mu_2 \nu_2}$ are uniquely determined as follows,
\begin{align}
N^{(4)\mu_1 \nu_1 \mu_2 \nu_2} := -M^{(4)(\mu_1 \nu_1 \mu_2 \nu_2)}
+\left(\frac12S^{(2)\alpha \beta (\mu_1 \nu_1} - N^{(2)\alpha \beta (\mu_1 \nu_1}\right) R^{\mu_2}_\alpha R^{\nu_2)}_\beta 
-\frac{1}{8} g^{\alpha \beta (\mu_1 \nu_1} (R^2)^{\mu_2}_\alpha (R^2)^{\nu_2)}_\beta. 
\end{align}

\section{Comparison with dRGT theory}
\label{cdr}
Restricting our argument to the leading order nonminimal coupling terms, let us compare our result with the linearized dRGT model given in (\ref{int3}).\footnote{The derivation of the action (\ref{int3}) is give in Appendix \ref{Ldr}.} Although the linearized dRGT model is not perfectly equivalent to our result, we will see that a trivial extension of the linearized dRGT model is perfectly equivalent to our result. 

By using Eqs.(\ref{yui2}), (\ref{kana3}), (\ref{nana11}), and (\ref{nanana7}), our resulting action can be expressed as follows,
\begin{align}
S= \int d^Dx \sqrt{-g} &\left[ \frac12 g^{\mu_1 \nu_1 \mu_2 \nu_2 \mu_3 \nu_3}\nabla_{\mu_1} h_{\mu_2 \nu_2 } \nabla_{\nu_1} h_{\mu_3 \nu_3}  
+\frac{1}{2} \left\{ m^2 g^{\mu_1 \nu_1 \mu_2 \nu_2} 
+ \gamma_1 R g^{\mu_1 \nu_1 \mu_2 \nu_2} \right. \right. \notag \\
 &\left. \left.+ \frac{\gamma_2}{2} \left(R^{\mu_1 [\nu_1}g^{ \nu_2] \mu_2} - R^{\mu_2 [\nu_1} g^{\nu_2]  \mu_1} \right) 
+R^{\mu_1 \mu_2 \nu_1 \nu_2}\right\} h_{\mu_1 \nu_1} h_{\mu_2 \nu_2}+ \mathcal{O}\left(R^2/m^2 \right) \right]. \label{res1}
\end{align}
This action is different from the action (\ref{int1}) by the free parameter $\gamma_3$.

We find that our result (\ref{res1}) includes certainly the linearized dRGT model (\ref{int3}) by tuning the free parameter as follows,
\begin{align}
 \gamma_1=\frac{ s_2D -1}{2(D-1)}, \ \ \gamma_2=-4s_2.
\end{align}
This is natural result. Although we assumed the nonminimal coupling terms without derivatives, the linearized dRGT model does not contain any derivatives in the nonminimal coupling terms.\footnote{By considering the derivation of the action (\ref{int3}), we find that the linearized dRGT model does not depend on the derivative nonminimal coupling terms. Furthermore, we find that the terms depending on the free parameters depend only on the Ricci curvatures and the scalar curvatures.}
Although it seems that one more free parameter have not been constrained,
there ought to be no more constraints on the leading order nonminimal coupling terms because there is an extension of the linearized dRGT model.

The extension is almost trivial. 
Let us consider a model replaced the mass parameter $m^2$ with a local scalar function $\mu^2(x)$,
\begin{align}
S=\int d^Dx \sqrt{-g} \left[\frac12 g^{\mu_1 \nu_1 \mu_2 \nu_2 \mu_3 \nu_3}\nabla_{\mu_1} h_{\mu_2 \nu_2 } \nabla_{\nu_1} h_{\mu_3 \nu_3}
+\frac{1}{2} \left\{ \mu^2(x) g^{\mu_1 \nu_1 \mu_2 \nu_2} +{T'}^{\mu_1 \nu_1 \mu_2 \nu_2} +{N'}^{\mu_1 \nu_1 \mu_2 \nu_2} \right\}
h_{\mu_1 \nu_1} h_{\mu_2 \nu_2}\right]. \label{niini1}
\end{align}
Reconsidering the derivation of the condition (\ref{yua3}),
the condition corresponding to the action (\ref{niini1}) is given by,
\begin{align}
&\text{Det}({V'}^{0\mu\nu0}g_{\nu\rho}):=0,\notag \\
&{V'}^{0\mu\nu0}= -\frac{\mu^2(x)}{2} g^{00} \theta^{\mu\nu} +  \bar{S'}^{0\mu \nu 0} + {N'}^{0\mu \nu 0}
+Q^{0\mu \nu 0}, \notag \\
&\bar{S'}^{\mu\nu\mu_1 \nu_1} \equiv  {S'}^{(\mu\nu)(\mu_1 \nu_1)}, \notag \\
&{S'}^{\mu\nu\mu_1 \nu_1} \equiv {T'}^{\mu\nu\mu_1 \nu_1} + \delta T^{\mu\nu\mu_1 \nu_1}. \label{niini2}
\end{align}
Here, $Q^{\mu_1 \nu_1 \mu_2 \nu_2},\delta T^{\mu_1 \nu_1\mu_2 \nu_2}$ are the same ones defined in Eqs.(\ref{nana11}).
Although the terms with covariant derivative of $\mu^2(x)$ appear in the equation corresponding to Eq.(\ref{nana13}),
we should note that the terms with covariant derivative of $\mu^2(x)$ can be included into the neglected terms. 
Eq.(\ref{niini2}) is the condition just replaced $m^2$ with $\mu^2(x)$ in Eq.(\ref{yua3}).
Therefore, once we have obtained a solution $N^{\mu_1 \nu_1 \mu_2 \nu_2}(m^2), \ S^{\mu_1 \nu_1 \mu_2 \nu_2}(m^2)$ satisfying the condition (\ref{yua3}) in the full-order,
the model with tensors $N^{\mu_1 \nu_1 \mu_2 \nu_2}(\mu^2(x)), \ S^{\mu_1 \nu_1 \mu_2 \nu_2}(\mu^2(x))$ is also ghost-free for any scalar function $\mu^2(x)$.
Then, we can extend the linearized dRGT model (\ref{int3}) by replacing $m^2$ with arbitrary scalar function $\mu^2(x)$,\footnote{The dRGT model (\ref{int2}) can also be extend as the same way.
Indeed, the model replaced all the parameters $\beta_n$ with arbitrary local functions $\beta_n(x)$, which cannot depend on the dynamical metric $g_{\mu\nu}$ but can be depend on some external scalar fields, is also ghost-free as proved in \cite{generalizeddRGT}.
Although we can replace the free parameter $s_2$ in (\ref{niini3}) with an arbitrary local function, this replacement is not necessary for the argument in the leading order.}
\begin{align}
{S'}_{\text{dRGT}}\equiv&\int d^Dx \sqrt{-g} \left[  \frac12 g^{\mu_1 \nu_1 \mu_2 \nu_2 \mu_3 \nu_3}\nabla_{\mu_1} h_{\mu_2 \nu_2 } \nabla_{\nu_1} h_{\mu_3 \nu_3}  
+\frac{1}{2} \left\{ \mu^2(x) g^{\mu_1 \nu_1 \mu_2 \nu_2} 
+\frac{s_2D -1}{2(D-1)} Rg^{\mu_1 \nu_1\mu_2 \nu_2} \right. \right. \notag \\
 &\left. \left.- 2s_2\left(R^{\mu_1 [\nu_1}g^{ \nu_2] \mu_2} - R^{\mu_2 [\nu_1} g^{\nu_2]  \mu_1} \right) 
+  R^{\mu_1 \mu_2 \nu_1 \nu_2}\right\} h_{\mu_1 \nu_1} h_{\mu_2 \nu_2}+ \mathcal{O}\left(R^2/\mu^2(x) \right) \right].\label{niini3}
\end{align}
We can choose $\mu^2(x)$ as an arbitrary function constructed by background metric and curvature as follows,
\begin{align}
\mu^2(x) = m^2 + \alpha R + \mathcal{O}(R^2/m^2).\label{niini4}
\end{align}
Here, $\alpha$ is a new free parameter. Therefore, by substituting the above expression into the action (\ref{niini3}), we obtain,
\begin{align}
 {S'}_{\text{dRGT}}=&\int d^Dx \sqrt{-g} \left[  \frac12 g^{\mu_1 \nu_1 \mu_2 \nu_2 \mu_3 \nu_3}\nabla_{\mu_1} h_{\mu_2 \nu_2 } \nabla_{\nu_1} h_{\mu_3 \nu_3}  
+\frac{1}{2} \left\{ {m}^2 g^{\mu_1 \nu_1 \mu_2 \nu_2} 
+{\gamma'}_1 R g^{\mu_1 \nu_1\mu_2 \nu_2} \right. \right. \notag \\
 &\left. \left. +\frac{{\gamma'}_2}{2}\left(R^{\mu_1 [\nu_1}g^{ \nu_2] \mu_2} - R^{\mu_2 [\nu_1} g^{\nu_2]  \mu_1} \right) 
+  R^{\mu_1 \mu_2 \nu_1 \nu_2}\right\} h_{\mu_1 \nu_1} h_{\mu_2 \nu_2}+ \mathcal{O}\left(R^2/m^2 \right) \right]. \notag \\
 {\gamma'}_1 \equiv& \alpha + \frac{s_2D -1}{2(D-1)}, \ \ \ {\gamma'}_2 \equiv  -4s_2. \label{niini5}
\end{align}
We should note that the parameters ${\gamma'}_1, {\gamma'}_2$ are no longer related with each other due to the new free parameter $\alpha$ defined in Eq.(\ref{niini4}).
Hence, we can regard both of the parameters ${\gamma'}_1,{\gamma'}_2$ as free parameters.
Therefore, in the leading order, our resulting action (\ref{res1}) is perfectly equivalent to the trivial extension of the linearized dRGT model (\ref{niini5}).

\if0
In the previous section, we have confirmed the fact that the lower order nonminimal coupling terms are constrained by the higher order conditions.
Thus, the remaining two parameters may be constrained by the fifth or higher order conditions.    
Using the identities given in Sec.\ref{lieder}, it seems that we can solve the fifth or higher order conditions without hardships. 
\fi

\section{SUMMARY}

In this paper, we have investigated the linear theory describing the massive spin-two field which holds the ghost-freeness for any background metric.
Our only assumption is that the nonminimal coupling terms do not contain any derivatives acting on the massive spin-two fields.
We show that any nonminimal coupling terms can be decomposed into the terms expressed by the totally symmetric tensor $N^{\mu_1 \nu_1 \mu_2 \nu_2}$ and the mixed symmetric tensor $T^{\mu_1 \nu_1\mu_2 \nu_2}$ (or $S^{\mu_1 \nu_1 \mu_2 \nu_2}$).
By solving the condition for the ghost-freeness perturbatively, we obtain the constraints on the nonminimal coupling terms. 

The results on the detailed structures can be summarized as follows,
\begin{itemize}
\item
We confirm explicitly that the nonminimal coupling terms cannot be truncated at the leading order.
The second or higher order nonminimal coupling terms are necessary for the ghost-freeness.  
\item 
We find that the totally symmetric tensor $N^{\mu_1 \nu_1 \mu_2 \nu_2}$ is uniquely determined as some functions of the mixed symmetric tensor $T^{\mu_1 \nu_1mu_2 \nu_2}$ (or $S^{\mu_1 \nu_1 \mu_2 \nu_2}$). Then, all the ambiguity of the theory belong to the mixed symmetric tensor.
\item 
Furthermore, we obtain an additional constraint on the leading
order mixed symmetric tensor $T^{(1)\mu_1 \nu_1 \mu_2 \nu_2}$ (or $S^{(1)\mu_1 \nu_1 \mu_2 \nu_2}$) by solving the fourth order condition. The resulting nonminimal coupling terms in the leading order contain two free parameters.
\end{itemize}

A purpose of this paper is to explain the discrepancy between the bottom-up result in leading order and the linearized dRGT model.
As the result, we succeed to reduce the number of the free parameters in the leading order nonminimal coupling terms by considering the fourth order condition, and we confirm that the leading order nonminimal coupling terms of a trivial extension of the linearized dRGT model is perfectly equivalent to our resulting nonminimal coupling terms in the leading order.

If we treat the condition non-perturbatively, we can perhaps show the uniqueness of the trivial extension of the linearized dRGT model.\footnote{Here, the ``trivial extension of the linearized dRGT model" can be defined by replacing all the parameters $\beta_n$ in the linearized dRGT model (\ref{int3}) with arbitrary scalar function of the background metric and the partial derivatives of the metric.}
It may lead us to non-perturvative interpretation to compare our method with the linearized dRGT model given in \cite{bernard1}.
Unfortunately, the explicit form of the linearized theory derived in \cite{bernard1} is not so simple.
However, it may be possible to read off the crucial structure for ghost-freeness without giving the explicit form of the linearized theory.
Indeed, in the case of the vielbein description of the dRGT model (for example, see  \cite{viel1,viel3}),
the constraints of the linearized model have obtained without giving the specific form of the linearized theory \cite{mazuet1,mazuet2}.
The metric formulation version may be possible.

Finally, we would like to mention the possibility of the extension to the spin-three case.
The massive spin-thee theory in the arbitrary background have been investigated in \cite{fukuma}.
The interesting result in \cite{fukuma} is that there is a combination of the leading order nonminimal coupling terms consisting with leading order condition.
By extending our method to the case of spin-three, we may investigate whether or not the leading order nonminimal coupling terms are constrained by higher order condition.

\appendix

\section{Properties of $g^{\mu_1 \nu_1 \cdots \mu_n \nu_n}$}
\label{ap1}
\renewcommand{\theequation}{A.\arabic{equation}}
\setcounter{equation}{0}
In this appendix, we summarize the properties of the higher rank tensor $g^{\mu_1 \nu_1 \mu_2 \nu_2 \cdots \mu_n \nu_n}$.
These properties are valid for $\eta^{\mu_1 \nu_1 \cdots \mu_n \nu_n}$ in the case of the flat spacetime. 

{\bf Definition:} Let us define the higher rank tensor $g^{\mu_1 \nu_1 \mu_2 \nu_2 \cdots \mu_n \nu_n}$ as follows,
\begin{align}
g^{\mu_1 \nu_1 \mu_2 \nu_2 \cdots \mu_n \nu_n} &\equiv 
n! \delta^{\mu_1}_{[\rho_1} \delta^{\mu_2}_{\rho_2} \cdots \delta^{ \mu_n }_{\rho_n]}
g^{\rho_1\nu_1} g^{\rho_2 \nu_2} \cdots g^{\rho_n \nu_n}
\notag \\
 &=\frac{-1}{(D-n)!} E^{\mu_1 \mu_2 \cdots \mu_n \sigma_{n+1} \cdots \sigma_D } E^{\nu_1 \nu_2 \cdots \nu_n}_{~~~~~~~~~~\sigma_{n+1} \cdots \sigma_D }.  \label{Ap1}
\end{align}
Here, $E^{\mu_1 \nu_1 \cdots \mu_n \nu_n}$ is defined by using the Levi-Civita anti-symmetric tensor density $\epsilon^{\mu_1 \nu_1 \cdots \mu_n \nu_n}$ as follows,
\begin{align}
&E^{\mu_1 \mu_2 \cdots \mu_D} \equiv \frac{1}{\sqrt{-g}} \epsilon^{\mu_1 \mu_2 \cdots \mu_D} \notag \\
&\epsilon^{\mu_1 \mu_2 \cdots \mu_D}=
\begin{cases}
+1\text{~~~} (\mu_1 \mu_2 \cdots \mu_D) \text{ is an even permutation of } (0123\cdots ) \\
 -1\text{~~~} (\mu_1 \mu_2 \cdots \mu_D) \text{ is an odd permutation of } (0123\cdots ) \\
0 \; \mbox{~~~~~Otherwise}  
\end{cases}
\end{align}
The symmetries of $g^{\mu_1 \nu_1\mu_2 \nu_2 \cdots\mu_n \nu_n}$ can be summarized as follows, 
\begin{align}
&\mu_i \longleftrightarrow \mu_j \text{ :Anti-symmetric} \notag \\
&\nu_i \longleftrightarrow \nu_j \text{ :Anti-symmetric} \notag \\
&(\mu_i , \nu_i ) \longleftrightarrow (\mu_j , \nu_j ) \text{ :Symmetric} \notag \\
&\{ \mu_i \} \longleftrightarrow \{ \nu_i \} \text{ :Symmetric} \label{Appp1}
\end{align}

{\bf Contraction:}
By using the expression of the second line of Eq.(\ref{Ap1}),
The contraction of $g^{\mu_1 \nu_1 \cdots \mu_n \nu_n}$ with respect to superscripts $\mu_n,\nu_n$ is proportional to
$2(n-1)$ th rank tensor $g^{\mu_1 \nu_1 \mu_{n-1} \nu_{n-1}}$,
\begin{align}
\label{Ap2}
{g^{\mu_1 \nu_1 \cdots \mu_{n-1} \nu_{n-1} \mu_n }}_{\mu_n}= (D-n+1) g^{\mu_1 \nu_1 \cdots \mu_{n-1} \nu_{n-1} } .
\end{align}

{\bf Expansion:}
From the definition (\ref{Ap1}), 
the $2n$th-rank tensor can be expanded in products of the $2m$th-rank tensor and the $2(n-m)$th-rank tensor,
\begin{align}
g^{\mu_1 \nu_1 \cdots \mu_n \nu_n} &= \delta^{\nu_1 ~\nu_2 \cdots \nu_n}_{~\lambda _1~ \lambda_2 \cdots \lambda_n} 
g^{\mu_1 \lambda_1} \cdots g^{\mu_n \lambda_n} \notag \\
&= \delta^{\nu_1 ~\nu_2 \cdots \nu_n}_{~\lambda _1~ \lambda_2 \cdots \lambda_n} \frac{1}{m! (n-m)!}
 g^{\mu_1 \lambda_1 \cdots \mu_m \lambda_m} g^{\mu_{m+1} \lambda_{m+1} \cdots \mu_n \lambda_n }.    \label{Ap3}
\end{align}
A few examples are given by,
\begin{align}
g^{\mu_1 \nu_1 \mu_2 \nu_2 \mu_3 \nu_3} &= g^{\mu_1 \nu_1 } g^{\mu_2 \nu_2 \mu_3 \nu_3} + g^{\mu_1 \nu_2 } g^{\mu_2 \nu_3 \mu_3 \nu_1} 
+ g^{\mu_1 \nu_3} g^{\mu_2 \nu_1 \mu_3 \nu_2} , \notag \\
g^{\mu_1 \nu_1 \mu_2 \nu_2 \mu_3 \nu_3 \mu_4 \nu_4} &= g^{\mu_1 \nu_1 } g^{\mu_2 \nu_2 \mu_3 \nu_3 \mu_4 \nu_4} -
 g^{\mu_1 \nu_2 } g^{\mu_2 \nu_1 \mu_3 \nu_3 \mu_4 \nu_4} 
- g^{\mu_1 \nu_3} g^{\mu_2 \nu_2 \mu_3 \nu_1 \mu_4 \nu_4} - g^{\mu_1 \nu_4} g^{\mu_2 \nu_2 \mu_3 \nu_3 \mu_4 \nu_1}.
\end{align}

{\bf 1+(D-1) decomposition:}
 Following formula is also useful,
\begin{align}
g^{00\mu_1 \nu_1 \mu_2 \nu_2 \cdots \mu_n \nu_n }
= g^{00} \theta^{\mu_1 \nu_1 \mu_2 \nu_2 \cdots \mu_n \nu_n}.\label{klkl1}
\end{align}
Here, by using $\theta^{\mu\nu}$ which is the projection operator living in $D-1$ space orthogonal to time-like direction $g^0_\mu$, $\theta^{\mu_1 \nu_1 \mu_2 \nu_2 \cdots \mu_n \nu_n}$ is defined as follows,
\begin{align}
&\theta^{\mu_1 \nu_1 \mu_2 \nu_2 \cdots \mu_n \nu_n} \equiv n! \delta^{\mu_1}_{[\rho_1} \delta^{\mu_2}_{\rho_2} \cdots \delta^{\mu_n}_{\rho_n]} \theta^{\rho_1 \nu_1}
\theta^{\rho_2 \nu_2} \cdots \theta^{\rho_n \nu_n}, \notag \\
&\theta^{\mu\nu} \equiv g^{\mu\nu} - \frac{g^{0\mu} g^{0\nu}}{g^{00}}.
\end{align}

\section{On the kinetic term}
\label{ap2}
\renewcommand{\theequation}{B.\arabic{equation}}
\setcounter{equation}{0}
In this appendix, the notation of the kinetic term is summarized.

In contrast to the case of the flat spacetime, in the case of the curved spacetime, our kinetic term defined by using $g^{\mu_1 \nu_1 \mu_2 \nu_2 \mu_3 \nu_3}$ [defined in (\ref{aaa1})] is not equivalent to the common kinetic terms of the linearized Einstein-Hilbert action.  
By expand the tensor $g^{\mu_1 \nu_1 \mu_2 \nu_2 \mu_3 \nu_3}$,
our kinetic term can be expressed as,
\begin{align}
\mathcal{L}_{\text{p-l}}\equiv& \frac12 g^{\mu_1 \nu_1 \mu_2 \nu_2 \mu_3 \nu_3}\nabla_{\mu_1} h_{\mu_2 \nu_2 } \nabla_{\nu_1} h_{\mu_3 \nu_3}  \notag \\
=&\frac12\nabla_\rho h \nabla^\rho h +\frac12\left( \nabla^\rho h_{\rho \lambda} \nabla^\sigma h_{\sigma}^{~\lambda} + \nabla^{\alpha} h^{\beta \gamma} \nabla_\beta h_{\alpha \gamma}\right) \notag \\
&- \nabla^\rho h_{\rho \sigma} \nabla^\sigma h - \frac12\nabla^{\alpha }h^{\beta \gamma} \nabla_{\alpha} h_{\beta \gamma}.
\end{align}
On the other hand, the common kinetic terms of the linearized Einstein-Hilbert action
are given by,
\begin{align}
\mathcal{L}_{\text{E-H}}\equiv \frac12\nabla_\rho h \nabla^\rho h + \nabla^{\alpha} h^{\beta \gamma} \nabla_\beta h_{\alpha \gamma}- \nabla^\rho h_{\rho \sigma} \nabla^\sigma h - \frac12\nabla^{\alpha }h^{\beta \gamma} \nabla_{\alpha} h_{\beta \gamma}.
\end{align}
There is a difference by the orders of the covariant derivatives between the second terms of $\mathcal{L}_{\text{p-l}}$ and the second terms of $\mathcal{L}_{E-H}$.
The difference is given by,
\begin{align}
\mathcal{L}_{\text{p-l}}-\mathcal{L}_{\text{E-H}}&=\frac12\left( \nabla^\rho h_{\rho \lambda} \nabla^\sigma h_{\sigma}^{~\lambda} - \nabla^{\alpha} h^{\beta \gamma} \nabla_\beta h_{\alpha \gamma}\right) \notag \\
&=-\frac12h_{\rho \lambda} [\nabla^\rho, \nabla^\sigma] h_\sigma^{~\lambda}+(\text{total derivative}) \notag \\
&=-\frac12 h_{\rho \lambda} \left(- R^{\rho \tau}h_{\tau }^{~\lambda} +
R^{\rho \sigma \lambda \tau } h_{\sigma \tau}\right)+(\text{total derivative}).
\end{align}

In \cite{Buchbinder1}, $\mathcal{L}_{\text{E-H}}$ is used as kinetic term.
Their resulting action in the leading order is expressed as,
\begin{align}
S = \int d^D x \sqrt{-g}& \left[ \mathcal{L}_{\text{E-H}} +\frac{1}{2} \left\{ m^2 g^{\mu_1 \nu_1 \mu_2 \nu_2} 
+ \xi_1 R g^{\mu_1 \nu_1 \mu_2 \nu_2} + 2\xi_3 R^{\mu_1 \mu_2 \nu_1 \nu_2}\right\} h_{\mu_1 \nu_1} h_{\mu_2 \nu_2} \right. \notag \\
 &\left. +(\frac{1}{2}-\xi_2) R^{\alpha \beta}h_{\alpha \sigma} h_{\beta}^{~\sigma} + \xi_2 R^{\alpha \beta} h_{\alpha \beta} h   
+ \mathcal{O}\left(R^2/m^2 \right)  \right] \notag \\
= \int d^D x \sqrt{-g}& \left[ \mathcal{L}_{\text{p-l}} +\frac{1}{2} \left\{ m^2 g^{\mu_1 \nu_1 \mu_2 \nu_2} 
+ \xi_1 R g^{\mu_1 \nu_1 \mu_2 \nu_2}+ (2\xi_3+1) R^{\mu_1 \mu_2 \nu_1 \nu_2}\right\} h_{\mu_1 \nu_1} h_{\mu_2 \nu_2} \right. \notag \\
 &\left. -\xi_2 R^{\alpha \beta}h_{\alpha \sigma} h_{\beta}^{~\sigma} + \xi_2 R^{\alpha \beta} h_{\alpha \beta} h   
+ \mathcal{O}\left(R^2/m^2 \right)  \right].
\end{align}
This action is equivalent to (\ref{int1}) by 
\begin{align}
\gamma_1 =\xi_1, \ \ \gamma_2=4\xi_2 , \ \ \gamma_3=2\xi_3 +1.
\end{align}

\renewcommand{\theequation}{C.\arabic{equation}}
\setcounter{equation}{0}

\section{Linearized dRGT model}
\label{Ldr}
In this appendix, we would like to derive the most general nonminimal coupling terms of the linearized dRGT model given in (\ref{int3}).

\subsection{Background solution}
Let us start with derive the background solution of the dRGT model.
The derivation of the background solution is based on \cite{Hassan3}.\footnote{Although \cite{Hassan3} derive the algebraic solution of the Bimetric gravity \cite{Hassan1},
same method can be applied to the dRGT model.} The EoM of the action (\ref{int2}) is given by,
\begin{align}
E_{\mu\nu}\equiv G_{\mu\nu} + m^2 \sum_{n=0}^{D-1} \beta_n Y^{(n)}_{\mu\nu} (\mathcal{S})=0.\label{ldr1}
\end{align}
Here, $G^\mu_{~\nu}\equiv R^\mu_{~\nu} -Rg_{\mu\nu}/2$ denotes the Einstein tensor, $\mathcal{S}$ is the square root matrix defined in (\ref{int2}), and $Y^{(n)\mu}_{~~~~~~\nu}(\mathcal{S})$ is defined as follows,
\begin{align}
Y^{(n)\mu\nu}(\mathcal{S}) \equiv \frac{1}{n!} g^{\mu\nu\mu_1 \nu_1 \mu_2 \nu_2 \cdots \mu_n \nu_n} \mathcal{S}_{\mu_1 \nu_1}
\mathcal{S}_{\mu_2 \nu_2} \cdots \mathcal{S}_{\mu_n \nu_n}. 
\end{align}
In this section, we lower the indexes of $\mathcal{S}^\mu_{~\nu}$and $Y^{(n)\mu\nu}$ by using metric $g_{\mu\nu}$,
and raise the indexes by using the inverse matrix $g_{\mu\nu}$.

The EoM (\ref{int2}) can be solved for $\mathcal{S}^\mu_{~\nu}$ algebraically, because there are no derivative terms with respect to $\mathcal{S}^\mu_{~\nu}$.
Let us start with deriving the algebraic solution of the background equation (\ref{ldr1}).
We assume the solution of the matrix $\mathcal{S}^\mu_{~\nu}$ expanded in powers of curvatures $R/m^2$,
\begin{align}
&\mathcal{S}^\mu_{~\nu} =  \delta^\mu_\nu +M^\mu_{~\nu}, \notag \\
&M^{\mu}_{~\nu} \equiv \sum_{n=1}^{\infty} \frac{1}{m^{2n}}M^{(n)\mu}_{~~~~~\nu}. \label{ldr2}
\end{align}
Here, $M^{(n)\mu}_{~~~~~\nu}$ denotes the $n$th-order terms with respect to curvature. Then, $M^\mu_{~\nu}$ depends on the first or higher order terms with respect to curvature.
By using the formula given in (\ref{Ap2}), we can easily show the following relation,
\begin{align}
&\sum_{n=0}^{D-1} \beta_n Y^{(n)\mu\nu} (\mathcal{S})
= \sum_{k=0}^{D-1} s_k Y^{(k)\mu \nu} (M), \notag \\
&s_k \equiv \sum_{n=k}^{D-1} \beta_n  {}_{D-1-k}C_{n-k}. \label{ldr3}
\end{align}
We can choose $m^2,s_2,s_3,\cdots s_{D-1}$ as free parameters. Let us explain this fact.
By substituting the assumption (\ref{ldr2}) into the background equation (\ref{ldr1}) and using the above relation (\ref{ldr3}),
we obtain,
\begin{align}
G_{\mu\nu} +m^2s_0 g_{\mu\nu}
+ m^2 \sum_{k=1}^{D-1}s_k  Y^{(k)}_{\mu\nu} (M)=0.
\end{align}
The zeroth-oder equation,
\begin{align}
s_0 \equiv \sum_{n=0}^{D-1} \beta_n  {}_{D-1}C_{n} =0, \label{ldr4}
\end{align}
can be regarded as a constraint on the parameters $\beta_n$.\footnote{If we assume more general solution,
\begin{align}
\mathcal{S}^\mu_{~\nu} =  a\delta^\mu_\nu +aM^\mu_{~\nu}, \ \ a\text{ \ :constant} \label{abc2}
\end{align}
instead of the assumption (\ref{ldr2}),
Eq.(\ref{ldr4}) is rewritten as follows,
\begin{align}
s'_0\equiv \sum_{n=0}^{D-1} \beta_n a^n {}_{D-1}C_{n} =0.\label{abc1}
\end{align} 
We can regard (\ref{abc1}) as the condition constraining the factor $a$. 
Then, it seems that there are no constraints on the parameters $\beta_n$. 
However, the parameter $a$ appear in the resulting linearized equation only through the parameter ${s'}_k \equiv \sum_{n=k}^{D-1} \beta_n a^n  {}_{D-1-k}C_{n-k}  \ (k=1,2,\cdots D-1)$.
Hence, $a$ cannot be regarded as a free parameter independently of ${s'}_k\  (k=1,2,\cdots D-1)$ in the linearized equation, and we obtain a linearized equation just obtained by replacing $s_k$ in the resulting linearized equation under the assumption (\ref{ldr2}) with ${s'}_k$ defined  above. Therefore, it is not necessary to adopt the assumption (\ref{abc1}).}
Furthermore, there is a ambiguity of the choice of the free parameters because the mass parameter $m^2$ is just an overall factor of the linear combination of the other parameters $\beta_n$ in the background EoM (\ref{ldr1}).
We can fix this ambiguity by constraining $s_1$ as,
\begin{align}
s_1 \equiv \sum_{n=1}^{D-1} \beta_n {}_{D-2}C_{n-1} =1.
\end{align}
This condition makes the mass parameter $m^2$ the Fierz-Pauli mass in the linearized equation as we will see later.
Therefore, in our notation, the free parameters of the linearized dRGT model are expressed by $m^2, s_2, s_3 \cdots s_{D-1}$.

In the leading order of Eq.(\ref{ldr3}), we obtain the equation, 
\begin{align}
G_{\mu\nu}+g_{(\mu\nu)}^{~~~~\mu_1 \nu_1} M^{(1)}_{\mu_1 \nu_1} =0. \label{ldr5}
\end{align}
If we define the inverse matrix of the tensor $g_{(\mu\nu)}^{~~~~\mu_1 \nu_1}$ as,
\begin{align}
&{g^{-1}}_{\alpha \beta}^{~~~~,\mu\nu} g_{(\mu\nu)}^{~~~~\mu_1 \nu_1} 
= \delta^{\mu_1}_{(\alpha} \delta_{\beta)}^{\nu_1} ,\notag \\
&{g^{-1}}_{\alpha \beta}^{~~~~,\mu\nu}
\equiv \frac{g_{\alpha \beta}g^{\mu\nu}}{D-1} - \delta^{\mu}_{(\alpha} \delta_{\beta)}^{\nu},
\end{align}
we can easily solve the equation (\ref{ldr5}) as follows,
\begin{align}
&M^{(1)}_{\mu\nu} = P_{\mu\nu}, \notag \\
&P_{\alpha \beta } \equiv - {g^{-1}}_{\alpha \beta}^{~~~~,\mu\nu} G_{\mu \nu}
= R_{\alpha \beta} -\frac{R}{2(D-1)} g_{\alpha \beta}.\label{nono1}
\end{align}
The tensor $P_{\mu\nu}$ is called the Shouten tensor.

\subsection{Linearized equation}
{\bf Perturbative equation:} The linearization of Eq.(\ref{ldr1}) is given by,
\begin{align}
\delta E_\mu^{~\nu} &= \delta G_{\mu}^{~\nu} + m^2 
\sum_{n=0}^{D-1} \frac{1}{n!}\beta_n \delta \left[ \delta_{\mu~~\mu_1~~\mu_2 \cdots\mu_n}^{~~\nu~~~\nu_1~~\nu_2 \cdots \nu_n} \mathcal{S}^{\mu_1}_{~~\nu_1} \mathcal{S}^{\mu_2}_{~~\nu_2} \cdots \mathcal{S}^{\mu_n}_{~~\nu_n}  \right] \notag \\
&=
\delta G_{\mu}^{~\nu} + m^2 
\sum_{n=1}^{D-1} \beta_n Y^{(n-1)~~\nu ~~~\nu_1}_{~~~~~~~\mu~~,\mu_1}(\mathcal{S}) \delta \mathcal{S}^{\mu_1}_{~~\nu_1} , \notag \\
Y^{(n-1)~~\nu ~~~\nu_1}_{~~~~~~~\mu~~,\mu_1}(\mathcal{S}) &\equiv \frac{1}{(n-1)!}
\delta_{\mu~~\mu_1~~\mu_2 \cdots\mu_n}^{~~\nu~~~\nu_1~~\nu_2 \cdots \nu_n} \mathcal{S}^{\mu_2}_{~~\nu_2} \cdots \mathcal{S}^{\mu_n}_{~~\nu_n}.
\end{align}
By substituting Eq.(\ref{ldr2}) into $Y^{(n-1)~~\nu ~~~\nu_1}_{~~~~~~~\mu~~,\mu_1}(\mathcal{S})$, 
we obtain,
\begin{align}
\delta E_\mu^{~\nu} &= \delta G_{\mu}^{~\nu} + m^2 
\sum_{k=1}^{D-1} s_k 
Y^{(k-1)~~\nu ~~~\nu_1}_{~~~~~~~\mu~~,\mu_1}(M)
\delta \mathcal{S}^{\mu_1}_{~~\nu_1}.\label{lldr1}
\end{align}
{\bf Einstein tensor:} The perturbation of the Einstein tensor $G^\mu_{~\nu}$ is given by,
\begin{align}
\delta G^\rho_{~\nu} =  (\delta g^{\rho \mu}) R_{\mu \nu}+g^{\rho \mu} \delta R_{\mu \nu}  
- \frac{\delta^\rho_\nu}{2} (\delta g^{\alpha \beta}) R_{\alpha \beta} 
- \frac{\delta^\rho_\nu}{2} g^{\alpha \beta}\delta R_{\alpha \beta}.
\end{align}
The summation of the second term and the fourth term is expressed as,
\begin{align}
g^{\rho \mu} \delta R_{\mu \nu}  
- \frac{\delta^\rho_\nu}{2} g^{\alpha \beta}\delta R_{\alpha \beta}
&=\frac12 g^{\rho \mu}\left[ 2\nabla^\sigma \nabla_{(\mu} h_{\nu) \sigma} - \Box h_{\mu\nu} - \nabla_\mu \nabla_\nu h - \nabla^\alpha \nabla^\beta h_{\alpha \beta} g_{\mu \nu} + \Box h g_{\mu \nu}  \right] \notag \\
&=\frac12 g^{\rho \mu}\left[ g_{(\mu\nu)}^{~~~~(\mu_1 \nu_1) \mu_2 \nu_2} \nabla_{\mu_2} \nabla_{\nu_2}  +  R^{(\mu_1~~~~\nu_1)}_{~~~(\mu\nu)} 
+ R_{(\mu}^{~(\mu_1} \delta_{\nu)}^{\nu_1)} \right]h_{\mu_1 \nu_1}. 
\end{align}
Here, we define the perturbation of the metric $\delta g_{\mu\nu} \equiv h_{\mu\nu}$.
Then, we obtain,
\begin{align}
 \delta G^\rho_{~\nu} = \frac12 g^{\rho \mu}\left[ g_{(\mu\nu)}^{~~~~(\mu_1 \nu_1) \mu_2 \nu_2} \nabla_{\mu_2} \nabla_{\nu_2}  +  R^{(\mu_1~~~~\nu_1)}_{~~~(\mu\nu)} 
+ R_{(\mu}^{~(\mu_1} \delta_{\nu)}^{\nu_1)} -2\delta_{\mu}^{(\mu_1}  R^{\nu_1)}_{\nu} 
+ g_{\mu\nu} R^{\mu_1 \nu_1} \right]h_{\mu_1 \nu_1}.\label{lldr3}
\end{align}

{\bf Square root matrix:}
From the definition of $\mathcal{S}$ given in (\ref{int2}), the perturbation $\delta \mathcal{S}$ obeys following relation,
\begin{align}
\delta \mathcal{S}^\mu_{~\rho} \mathcal{S}^\rho_{~\nu} +
\mathcal{S}^\mu_{~\rho} \delta \mathcal{S}^\rho_{~\nu} = - h^{\mu \rho} f_{\rho \nu}.
\end{align}
By substituting Eq.(\ref{ldr2}) and using $f_{\mu\nu} = g_{\mu\rho} \mathcal{S}^\rho_{~\sigma}\mathcal{S}^\sigma_{~\nu}$, in the leading order approximation with respect to $R/m^2$, we obtain,
\begin{align}
\delta \mathcal{S}^\mu_{~\nu} &= -\frac{1}{2} h^\mu_{~\nu} 
+ \frac{1}{4m^2} \left[M^{(1)\mu\sigma}h_{\sigma\nu} -3 h^{\mu\sigma}M^{(1)}_{\sigma \nu} \right]
+ \mathcal{O}\left(\frac{R^2}{m^4}\right) \notag \\
&= -\frac{1}{2} h^\mu_{~\nu} 
-\frac{1}{2m^2}  g^{\mu\rho}\left[M^{(1)\sigma}_{(\rho}h_{\nu)\sigma} +2 h^{\sigma}_{~[\rho}M^{(1)}_{\nu] \sigma } \right]+ \mathcal{O}\left(\frac{R^2}{m^4}\right).\label{lldr2}
\end{align}

{\bf Zeroth-order:}
By substituting (\ref{lldr2}) into (\ref{lldr1}), in the zeroth-order, we obtain the Fierz-Pauli mass term,
\begin{align}
\delta E_\mu^{~\nu} \Bigl|_{\text{zeroth}} &= 
- \frac{m^2}{2}
g^{~~\nu ~~\nu_1}_{\mu~~\mu_1} h^{\mu_1}_{~~\nu_1}.
\end{align}

{\bf Leading order:}
By substituting (\ref{lldr3}) and (\ref{lldr2}) into (\ref{lldr1}), in the leading order,
we obtain the kinetic term and the leading order nonminimal coupling terms,
\begin{align}
\delta E_\mu^{~\nu} \Bigl|_{\text{leading}} 
=& \frac12 g^{\nu\rho}\left[ g_{(\mu\rho)}^{~~~~(\mu_1 \nu_1) \mu_2 \nu_2} \nabla_{\mu_2} \nabla_{\nu_2}  +  R^{(\mu_1~~~~\nu_1)}_{~~~(\mu\rho)} 
+ R_{(\mu}^{~(\mu_1} \delta_{\rho)}^{\nu_1)} -2\delta_{\rho}^{(\mu_1}  R^{\nu_1)}_{\mu} 
+ g_{\mu\rho} R^{\mu_1 \nu_1} \right]h_{\mu_1 \nu_1} \notag \\
&- \frac{1}{2} g^{~~\nu \mu_1 \nu_1}_{\mu} 
  \left[M^{(1)\sigma}_{(\mu_1}h_{\nu_1)\sigma} +2 h^{\sigma}_{~[\mu_1}M^{(1)}_{\nu_1] \sigma } \right] - \frac{s_2}{2} g_{\mu}^{~~\nu\mu_1 \nu_1 \mu_2 \nu_2} M^{(1)}_{\mu_1 \nu_1} 
 h_{\mu_2\nu_2} \notag \\
=&\frac12 g^{\nu\rho}\left[ g_{(\mu\rho)}^{~~~~(\mu_1 \nu_1) \mu_2 \nu_2} \nabla_{\mu_2} \nabla_{\nu_2} 
+  R^{(\mu_1~~~~\nu_1)}_{~~~(\mu\rho)} 
+\frac{R}{2(D-1)}g_{\mu\rho}^{~~~\mu_1 \nu_1} 
- s_2 g_{\mu \rho}^{~~~\mu_1 \nu_1 \mu_2 \nu_2} P_{\mu_2 \nu_2} 
 \right]h_{\mu_1 \nu_1}.
\end{align}
In the second line, we substitute the explicit form of $M^{(1)\mu\nu}$ given in (\ref{nono1}).
By substituting the explicit form of $P_{\mu\nu}$ given in (\ref{nono1}) into remaining Shouten tensor, 
we obtain,
\begin{align}
g^{\mu\rho}\left[\delta E_\rho^{~\nu} \Bigl|_{\text{zeroth}}+
\delta E_\rho^{~\nu} \Bigl|_{\text{leading}} \right] = 
&\frac12 g^{(\mu \nu) \mu_1 \nu_1 \mu_2 \nu_2 }
\nabla_{\mu_1} \nabla_{\nu_1} h_{\mu_2 \nu_2 } 
-\frac{1}{2} \left\{ m^2 g^{\mu \nu\mu_1 \nu_1 } 
+\frac{s_2D -1}{2(D-1)} Rg^{\mu \nu \mu_1 \nu_1} \right. \notag \\
 &\left. - 2s_2\left(R^{\mu [\nu}g^{ \nu_1] \mu_1} - R^{\mu [\nu} g^{\nu_1]  \mu_1} \right) 
+  R^{\mu \mu_1 \nu \nu_1}\right\} h_{\mu_1 \nu_1}.
\end{align}

This is the EoM derived from the action (\ref{int3}).

\renewcommand{\theequation}{D.\arabic{equation}}
\setcounter{equation}{0}

\section{Einstein manifold case}
\label{ein}
In the case of Einstein manifold, the Ricci curvature satisfies the following relation,
\begin{align}
R_{\mu\nu} = \frac{R}{D}g_{\mu\nu}. \label{ein1}
\end{align}
Thus, the independent curvatures are scalar curvature $R$
and Weyl curvature $C^{\mu_1 \nu_1 \mu_2 \nu_2}$.
In this case, we have proposed in \cite{new curved} that the model with,
\begin{align}
 N^{\mu_1 \nu_1 \mu_2 \nu_2} =0, \ \ S^{\mu_1 \nu_1 \mu_2 \nu_2} \text{ :not restricted},
\label{ein4}
\end{align}
is a solution of the condition (\ref{yua3}).
Here, $S^{\mu_1 \nu_1 \mu_2 \nu_2}$ is the arbitrary mixed symmetric tensor
which constructed by scalar curvature, Weyl curvature, and the covariant derivatives of Weyl curvature. 

This statement can easily be shown.
By substituting Eq.(\ref{ein1}) into the definition of $Q^{\mu_1 \nu_1 \mu_2 \nu_2}$ in Eq.(\ref{nana11}), we obtain,
\begin{align}
Q^{\mu_1 \nu_1 \mu_2 \nu_2}=0. \label{ein2}
\end{align}
Hence the tensor $V^{0\mu\nu0}$ becomes,
\begin{align}
V^{0\mu\nu0} = m^2 g^{0\mu\nu0} + \bar{S}^{0\mu\nu0}.
\end{align}
Using the relation (\ref{saya5}), we find that the tensor $V^{0\mu\nu0}$ has a zero eigenvector $g^0_\nu$, i.e.,
\begin{align}
V^{0\mu\nu0} g^0_\mu =0.
\end{align}
Therefore, the condition (\ref{yua3}) is satisfied.

Conversely, by using the method proposed in this paper, we can show that the solution (\ref{ein4}) is the general solution of the condition (\ref{nara2}).
By substituting (\ref{ein2}) into the condition (\ref{nara2}),
we obtain,
\begin{align}
\hat{N}^{00} + \sum_{n=0}^{\infty} \left( \frac{2}{m^2g^{00}} \right)^{n+1} \hat{N}^0_{~\nu} \left[\theta( \hat{N}\theta + \hat{S} )^n\right]^{\nu\mu} \hat{N}_\mu^{~0}  :=0.
\end{align}
By expanding the above condition in powers of $R/m^2$,
the leading order condition is,
\begin{align}
N^{(1)0000}:=0. \label{ein5}
\end{align}
The lemma given in Sec.\ref{lem} can easily be extended to the case of the Einstein manifold.
Eq.(\ref{ein1}) can be regarded as a constraint on the background metric, $\psi_{\mu\nu} \equiv R_{\mu\nu}-\frac{R}{D}g_{\mu\nu}=0$.
Then,`` for any variation $\delta g_{\mu\nu}$", Eq.(\ref{lem2}) is rewritten as follows,
\begin{align}
D^{0000} (g_{\alpha \beta}+\delta g_{\alpha \beta}) - D^{0000} (g_{\alpha \beta})
= \lambda^{\mu\nu} \left[ \psi_{\mu\nu}(g_{\alpha \beta} +\delta g_{\alpha \beta}) 
- \psi_{\mu\nu}(g_{\alpha \beta}) \right].
\end{align}
Here, $\lambda^{\mu\nu}$ is the Lagrange multiplier.
However, if we set the variation $\delta g_{\mu\nu}$ as the Lie derivative $\mathcal{L}_{\xi} g_{\mu\nu}$, the variation of the constraint becomes equal to zero under the original equation $\psi_{\mu\nu}=0$,\footnote{This fact is due to the general covariance of the constraint $\psi_{\mu\nu}=0$.
For example, in the case of the non-covariant constraint $\psi \equiv R_{00}-\frac{R}{D}g_{00} =0$, the Lie derivative $\mathcal{L}_\xi \psi $ is not equal to zero under the constraint $\psi=0$.}
\begin{align}
\psi_{\mu\nu}(g_{\alpha \beta} +\delta g_{\alpha \beta}) 
- \psi_{\mu\nu}(g_{\alpha \beta})& = \mathcal{L}_\xi \psi_{\mu\nu} \notag \\
&= \xi^\alpha \partial_\alpha \psi_{\mu\nu} + 2 \partial_{(\mu} \xi^\rho \psi_{\nu)\rho}
 = 0,
\end{align}
Therefore, the lemma is also valid in the case of Einstein manifold.

Using the lemma, the condition (\ref{ein5}) is rewritten as follows,
\begin{align}
 N^{(1)\mu_1 \nu_1 \mu_2 \nu_2} :=0. \label{ein3}
\end{align}
By using the above result (\ref{ein3}), the second order condition becomes,
\begin{align}
N^{(2)0000}:=0 \longrightarrow N^{(2)\mu_1 \nu_1\mu_2 \nu_2} :=0.
\end{align}
As the same way, in any order, we obtain the condition,
\begin{align}
N^{(n)0000}:=0 \longrightarrow N^{(n)\mu_1 \nu_1 \mu_2 \nu_2}:=0.
\end{align}
Therefore, the solution (\ref{ein4}) is the general solution of the condition (\ref{yua3}).

\renewcommand{\theequation}{E.\arabic{equation}}
\setcounter{equation}{0}
\section{Comparison with String theory}
\label{str}
In \cite{Buchbinder1}, the effective field theory (EFT) of the bosonic string theory without any brains is investigated.
Although it had been pointed out that the EFT of the string theory with massive fields cannot become general covariant \cite{tseytlin}, Buchbinder {\it et al.} pick up the covariant parts by only consider the interaction between the massive spin-two mode in the open string and the massless graviton in the closed string.
They consider the open string coupled with the background gravity and background massive spin-two field.
By requiring the quantum Weyl invariance in the one-loop level,  
the equations of motion of the background fields are derived.
\footnote{Such a method is proposed in \cite{sigma1,sigma2,sigma3,sigma4} }
Although it seems not necessary that the EFT of the string theory
is included in the fixed curved background model\footnote{There is a possibility that the ghost-freeness of EFT of the string theory cannot be explained without using the background EoM of the gravity sector.
Furthermore, there is another possibility that the EFT contains higher derivative terms of massive spin-two field in contrast to our later assumption given in (\ref{stri}).},
it is interesting to compare our result with string theory.
The coupling terms between the gravity and the massive spin-two field of resulting EFT is included in the their leading order result (\ref{int1}),
and the parameter is given by,
\begin{align}
\gamma_1=0, \ \ \gamma_2= -2, \ \  \gamma_3=-1. \label{str1}
\end{align}
This tuning is incompatible to our result (\ref{res1}),
because our result gives $\gamma_3=1$.
We can raise two possibilities explaining this difference.
First, as we have mentioned, there is possibility that the string theory cannot be understood from the point of view of the massive spin-two theory in the arbitrary background. 
However, there remains a possibility that the parameter region (\ref{str1}) may be allowed as the massive spin-two model in arbitrary background.
The counter terms of the string theory ought to contain some derivatives of $h_{\mu\nu}$ generally, 
because the calculation is just like as renormalization of the nonlinear sigma model.
Thus, the derivative nonminimal coupling terms may appear in the EFT of the string theory.
Our result (\ref{res1}) is derived under the assumption that the nonminimal coupling terms do not contain any derivatives acting on $h_{\mu\nu}$.
Therefore, the parameter region (\ref{str1}) may be allowed by considering the derivative nonminimal coupling terms.

The extension is not so difficult,
because we can restrict the forms of derivative nonminimal coupling terms from the beginning.
The general derivative nonminimal coupling terms violate the constraints corresponding to  
(\ref{nana8}) and (\ref{nana13}).
The derivative nonminimal coupling terms which do not violate the constraint (\ref{nana8}) and (\ref{nana13}) is given by,
\begin{align}
S_{D} = \int d^Dx \sqrt{-g} C^{\mu_1 \nu_1 \mu_2 \nu_2 \mu_3 \nu_3} \nabla_{\mu_1}h_{\mu_2 \nu_2} \nabla_{\nu_1}h_{\mu_3 \nu_3}. \label{stri}
\end{align}
Here, the sixth order tensor $C^{\mu_1 \nu_1 \mu_2 \nu_2 \mu_3 \nu_3}$
is the non-derivative tensor whose superscripts are anti-symmetric for the permutation of $\mu_i \leftrightarrow \mu_j$, $\nu_i\leftrightarrow \nu_j$ $(i,j=1,2,3 \ \text{and} \  i \neq j)$, that is 
just same symmetry as $g^{\mu_1 \nu_1 \mu_2 \nu_2 \mu_3 \nu_3}$.
Under the assumption, the constraints (\ref{nana8}) and (\ref{nana13}) exist.
Then, we can analyze the theory as the same way in the case of the nonderivative nonminimal coupling terms.

\end{document}